\documentclass[12pt]{iopart}
\usepackage{graphicx}
\usepackage{dcolumn}
\usepackage{bm}
\usepackage{epsf}
\usepackage{amssymb}
\usepackage{subfigure}
\usepackage{cancel}
\DeclareGraphicsExtensions{.pdf}
\begin{document}

\title[Joint Entanglement of Topology and Polarization]{Joint Entanglement of Topology and Polarization Enables Error-Protected Quantum Registers}

\author{David S. Simon$^{1,2}$, Shuto Osawa$^2$, and Alexander V. Sergienko$^{2,3,4}$}

\address{1. Dept. of Physics and Astronomy, Stonehill College, 320 Washington Street, Easton, MA 02357, USA}

\address{2. Department of Electrical and Computer Engineering, Boston
University, 8 Saint Mary's St., Boston, MA 02215, USA}

\address{3. Photonics Center, Boston
University, 8 Saint Mary's St., Boston, MA 02215, USA}

\address{4. Dept. of Physics, Boston University, 590 Commonwealth
Ave., Boston, MA 02215, USA}

\eads{simond@bu.edu, sosawa@bu.edu, alexserg@bu.edu}

\begin{abstract}
Linear-optical systems can implement photonic quantum walks that simulate systems with nontrivial topological properties. Here, such photonic walks are used to
jointly entangle polarization and winding number. This joint entanglement allows information processing tasks to be performed with interactive access to a wide
variety of topological features. Topological considerations are used to suppress errors, with polarization allowing easy measurement and manipulation of qubits.
We provide three examples of this approach: production of two-photon systems with entangled winding number (including topological analogs of Bell states), a
topologically error-protected optical memory register, and production of entangled topologically-protected boundary states. In particular it is shown that a pair
of quantum memory registers, entangled in polarization and winding number, with topologically-assisted error suppression can be made with qubits stored in
superpositions of winding numbers; as a result, information processing with winding number-based qubits is a viable possibility.
\end{abstract}

\noindent{\it Keywords\/}: quantum simulation, topologically protected states, linear optics, optical multiports

\submitto{\NJP} \maketitle
\maketitle

%


\section{Introduction}

States with integer-valued topological invariants, such as winding and Chern
numbers, exhibit a variety of physically interesting effects in solid-state
systems \cite{hasan,kitrev,asb,bern,stan}, including integer and fractional
quantum Hall effects \cite{klitz,laugh1,thoul,tsui,laugh2}. When systems with
different values of topological invariants are brought into contact, states
arise that are highly localized at the boundaries. These edge or boundary
states have unusual properties; for example, in two-dimensional materials
they lead to unidirectional conduction at the edges, while the interior bulk
remains insulating. Because of the inability to continuously interpolate
between discrete values of the topological invariant, these surface states
are protected from scattering and are highly robust against degradation,
making them prime candidates for use in error-protected quantum information
processing.

Optical states with similar topological properties can be produced by means
of photonic quantum walks in linear-optical systems
\cite{kitrev,bouw,zhang,souto,peretz,sch,broome,kit1,kit2,kit3,lu1,lu2,cardano,simbook}.
Photonic walks have demonstrated topological protection of
polarization-entanglement \cite{moul} and of path entanglement in photonic
crystals \cite{recht}.

Optical systems are useful laboratories to study topologically-nontrivial
states, due to the high level of control: system properties can be varied
over a wide range of parameters, in ways not easy to replicate in condensed
matter systems. In the quest to carry out practical quantum information
processing tasks, it is of great interest to examine more closely the types
of topologically-protected states that can be optically engineered. Those
that also entangled are of particular interest for quantum information
applications.

The goal here is to entangle states associated with distinct topological sectors, and
to do so in a way \emph{that allows this entangled topology to be readily
available for information processing}. Specifically, linear optics will be
used to produce: (i)  entangled topologically-protected boundary states, (ii)
winding-number-entangled bulk states,  and (iii) an entangled pair of
error-protected memory registers. To create the states, a source of initial
polarization-entangled light is necessary, specifically type-II spontaneous
parametric down conversion (SPDC) in a $\chi^{(2)}$ nonlinear crystal. Taking
this initial state as given, all further processing requires only linear
optical elements.

Topological invariants characterize global properties of systems and cannot be easily distinguished by localized measurements. This difficulty in measurement
limits their use in many applications. That problem is solved here by linking topology to a more easily-measured variable, polarization. Polarization and winding
number will be tightly correlated (and in fact, jointly entangled with each other), but will serve distinct purposes: winding number provides stability against
perturbations, while polarization allows easy access and measurement.

We confine ourselves to one-dimensional systems. After reviewing directionally-unbiased optical multiports, it is shown how arrays of such multiports can produce
entangled pairs of bulk states associated with Hamiltonians of different winding number, as well as entangled pairs of topologically-protected edge states
localized near boundaries between regions of different topology.

A variation of the same idea then allows single qubits or entangled qubit pairs to be stored in a linear optical network as winding numbers.  Topological
protection of boundary states is well-known, but less widely recognized is the fact that bulk wavefunctions also have a degree of resistance to changes in
winding numbers \cite{simwave}. This effect is discussed in the appendix and will be used to reduce polarization-flip errors of qubits stored in the optical
register, greatly reducing the need for additional error correction.

\section{Directionally-unbiased multiports and topological states}

\begin{figure}
\centering
\includegraphics[totalheight=1.8in]{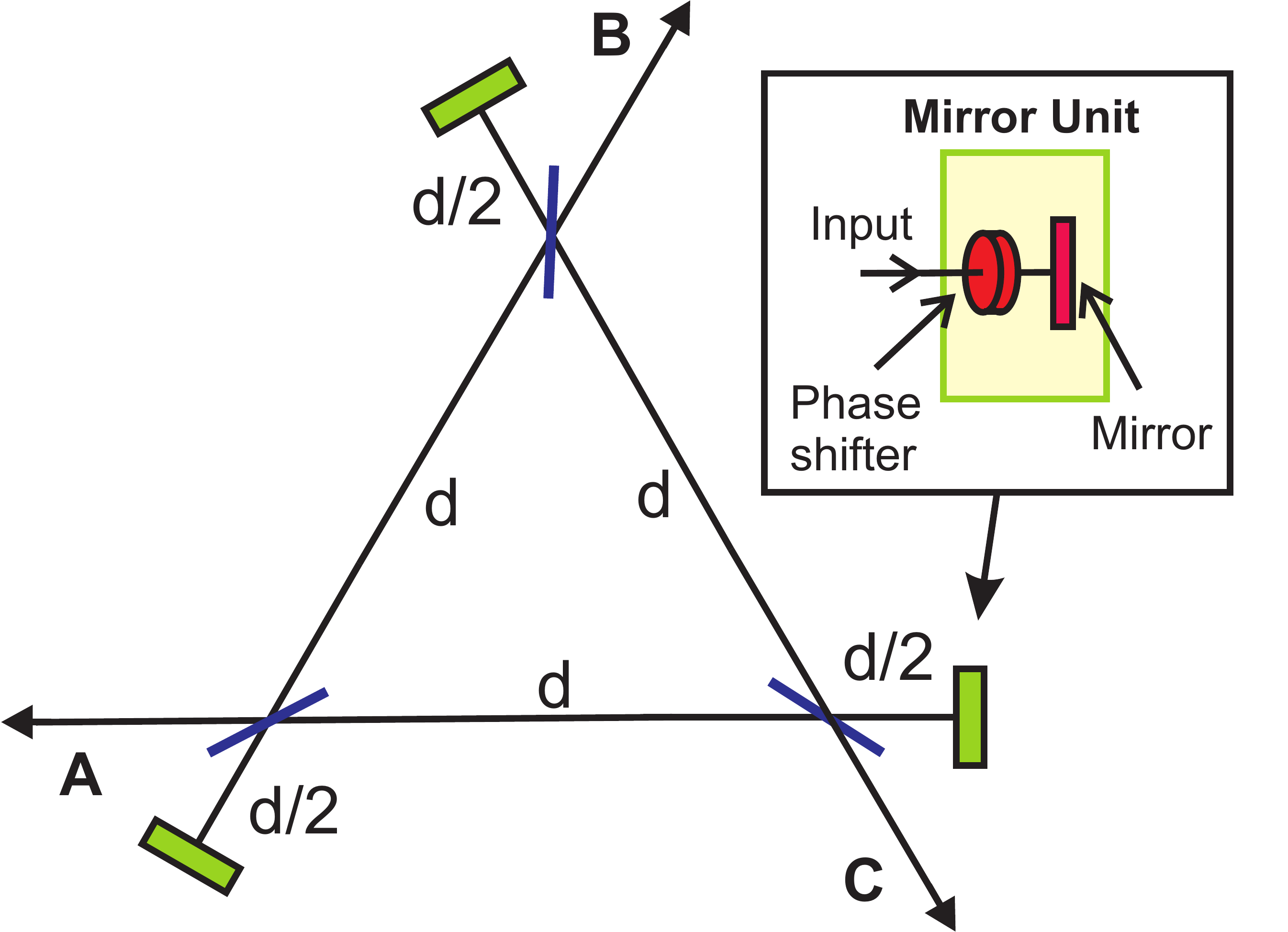}
\caption{Directionally-unbiased threeport unit: three beam
splitters form a triangular array with three external ports and three internal edges.
One port of each beam splitter passes through a phase shifter and then reflects
back onto itself via a mirror. Altering the phase shift imparted at these mirror
units allows control of the device properties by altering the interference
between paths. (Figure adapted from \cite{sim1}.) } \label{roundfig}
\end{figure}

In standard beam splitters and multiports, photons cannot reverse course to
exit back out the input port. In \cite{sim1}, a generalized multiport was
introduced which allows exit with some probability out any port, including
the input. These directionally-unbiased multiports have adjustable internal
parameters (reflectances and phase shifts) that allow the exit probabilities
at each port to be varied. The three-port version is shown schematically in Fig. \ref{roundfig}.

Directionally-unbiased multiports are linear optical devices with the input/output ports connected via beam splitters to vertex units of the form shown in the
inset of Fig. \ref{roundfig}. Each such unit contains a mirror and phase shifter.  The beam splitter-to-mirror distance ${d\over 2}$ is half of the distance
$d$ between the vertex units in the multiport. The phase shifter provides control of the properties of the multiport, since different choices of phase shift at
the vertices affect how the various photon paths through the device interfere with each other. These devices and some of their applications have been studied
theoretically in \cite{sim1,sim3,sim2} and experimentally demonstrated in \cite{osawa}.

If the unit is sufficiently small (quantitative estimates of the required
size and other parameter values may be found in \cite{sim1}) then its action can be described by
an $n\times n$ unitary transition matrix $\hat U$ whose rows and columns
correspond to the input and output states at the ports. If the internal phase
shifts at all the mirror units are known, then an explicit form of the
unitary transition matrix $\hat U$ can be given. Here, we restrict ourselve
to the three-port and assume that all three vertices are identical, in which
case \cite{sim3}:
\begin{equation}\hat U={{e^{i\theta}}\over {2+ie^{i\theta}}}\left( \begin{array}{ccc} 1 & ie^{-i\theta}-1 & ie^{-i\theta}-1 \\
ie^{-i\theta}-1 & 1 & ie^{-i\theta}-1\\ ie^{-i\theta}-1 & ie^{-i\theta}-1& 1
\end{array}\right) ,\label{general}\end{equation} where $\theta$ is the total phase shift at each mirror unit (including both the mirror and the phase plate).
The rows and columns refer to the three ports $A$, $B$, $C$.

\begin{figure}
\centering
\includegraphics[totalheight=1.3in]{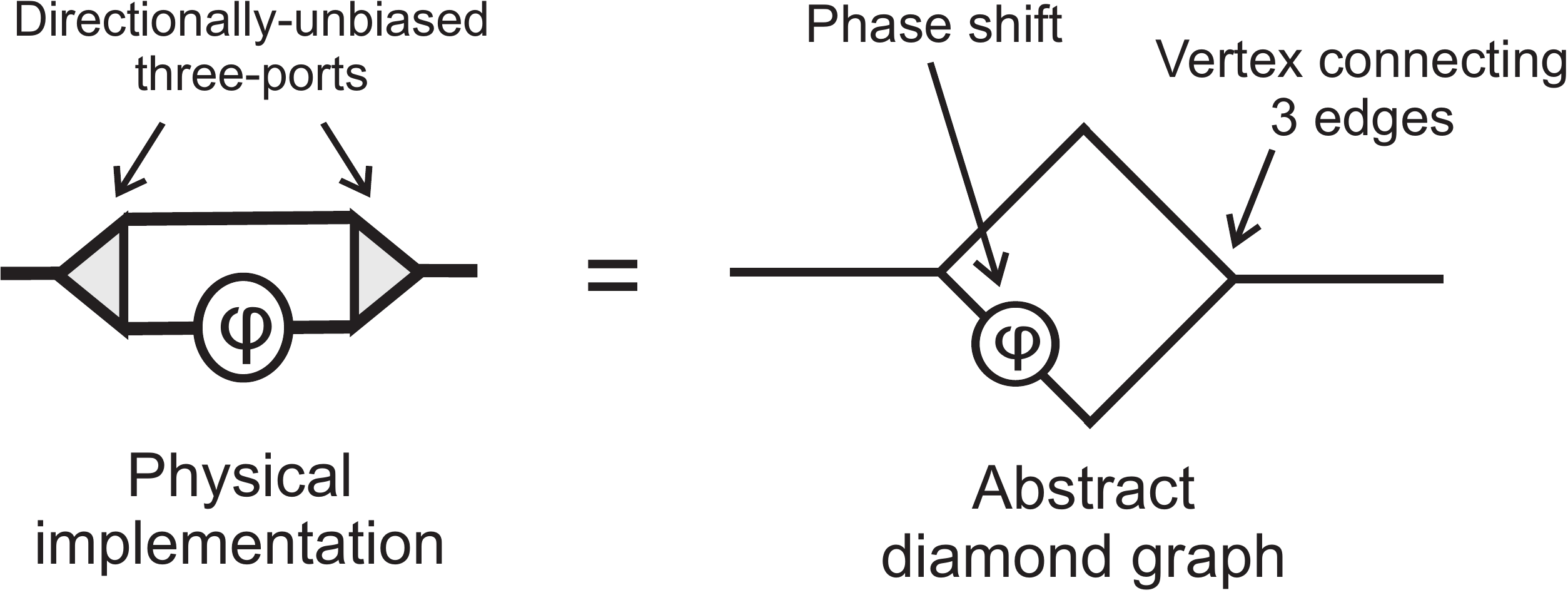}
\caption{Two multiports and a phase shifter are used to
construct a diamond-shaped structure. The multiports are viewed as scattering
centers at edges of an optical graph.
Detailed properties of this structure may be found in
\cite{sim3,feld1,feld2,feld3}. Alternating pairs of these diamond graphs with
different values of $\phi$ (see Fig. \ref{schemefig}) will be the
basic building blocks of the structures in the following sections.  (Figure reproduced from \cite{sim3}.)}
\label{diamondfig}
\end{figure}

Arrays of three-ports and phase shifters can simulate a range of
discrete-time Hamiltonian systems \cite{sim2}, including some with distinct
topological phases \cite{sim3}. The array acts as a lattice through which
photons propagate, leading to photonic analogs of Brillouin zones and energy
bands. By altering multiport parameters, systems can be simulated \cite{sim3}
in which Hamiltonians have different winding numbers as one wraps
around a full Brillouin zone.  As is well known from solid-state physics
\cite{hasan,asb,bern,stan}, at the boundaries between regions with distinct
winding number localized boundary or edge states appear. These edge states
are highly stable due to topological protection; different discrete winding
numbers on the two sides prevent the state from being destroyed by small,
continuous perturbations.

\begin{figure}
\centering
\includegraphics[totalheight=.6in]{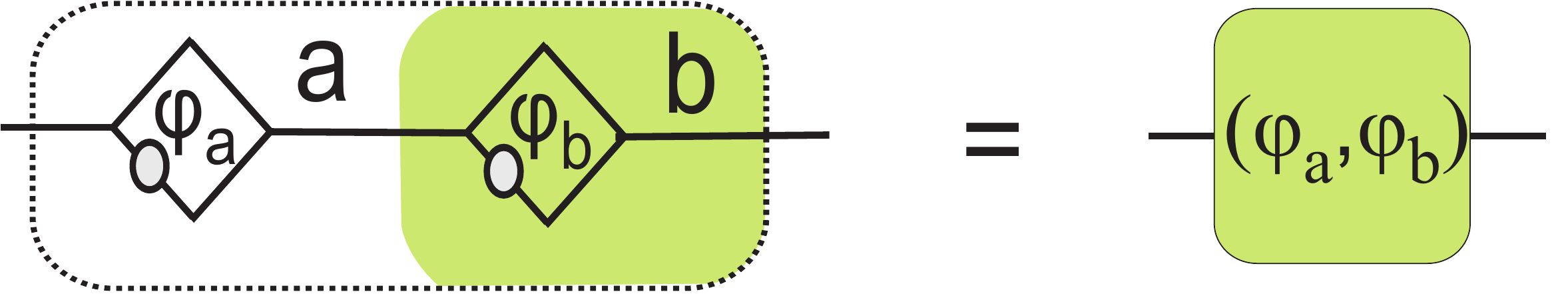}
\caption{The unit cell for the proposed systems is a pair of diamond graphs
with phase shifts $\phi_a$ (unshaded) and $\phi_b$ (shaded), as shown on the left. Each such cell
contains four three-ports.
This basic unit will
be drawn in the schematic form shown on the right. }
\label{schemefig}
\end{figure}

The basic building block for the optical systems described below is the diamond-graph structure \cite{sim3,feld1,feld2,feld3} formed by two unbiased three-ports
and an additional phase shifter. This is shown in Fig. \ref{diamondfig}, where the unbiased three-port is represented by a vertex connecting three edges. The
unit cell for the lattice structures will be formed by two such diamond graphs (Fig. \ref{schemefig}), and so each cell will contain four multiports. The phase
shifts in each of the two diamonds may be different, $\phi_a\ne \phi_b$. When a string of these unit cells are connected end-to-end, photons inserted into the
chain exhibit quantum walks. The resulting system is governed by a Hamiltonian which can have nontrivial topological structure: depending on the values of the
phases $\phi_a$ and $\phi_b$, the winding number $\nu$ can take a value of either $0$ or $1$ \cite{sim3}. When a chain of winding number $\nu =1$ is connected to
a chain with $\nu =0$, localized topologically-protected edge states appear at the boundary \cite{sim3}.

Additional discussion of topological aspects of one-dimensional models is given in the appendix, along with numerical simulations displaying these properties in
systems formed by chains of unbiased multiport.

\section{Jointly-entangled topologically-protected bulk
states}\label{hyperbulk}

Start with a polarization-entangled photon source, type-II spontaneous parametric down conversion in a nonlinear crystal. With appropriate filtering and phase
shifts, the two-photon output can be taken to be an entangled Bell state, \begin{equation}|\psi^\pm \rangle = {1\over\sqrt{2}}\left( |H\rangle_1 |V\rangle_2 \pm
|V\rangle_1 |H\rangle_2\right) ,\label{bell1}\end{equation} where $1$ and $2$ refer to two spatial modes. We wish now to create from this a state of entangled
winding number. Polarization entanglement should remain intact, to use for control and measurement purposes.

It should be noted that we are making a slight abuse of terminology here: the winding number is associated with the Hamiltonian, not strictly speaking with the
state. But as discussed in the appendix, transitions of bulk states between spatial regions or parameter values with Hamiltonians of different winding number can
be arranged to be strongly suppressed. Therefore, as a practical matter, under appropriate conditions one may to a high degree of accuracy think of the winding number as being associated
with the state as well.

\begin{figure}
\centering
\includegraphics[totalheight=1.5in]{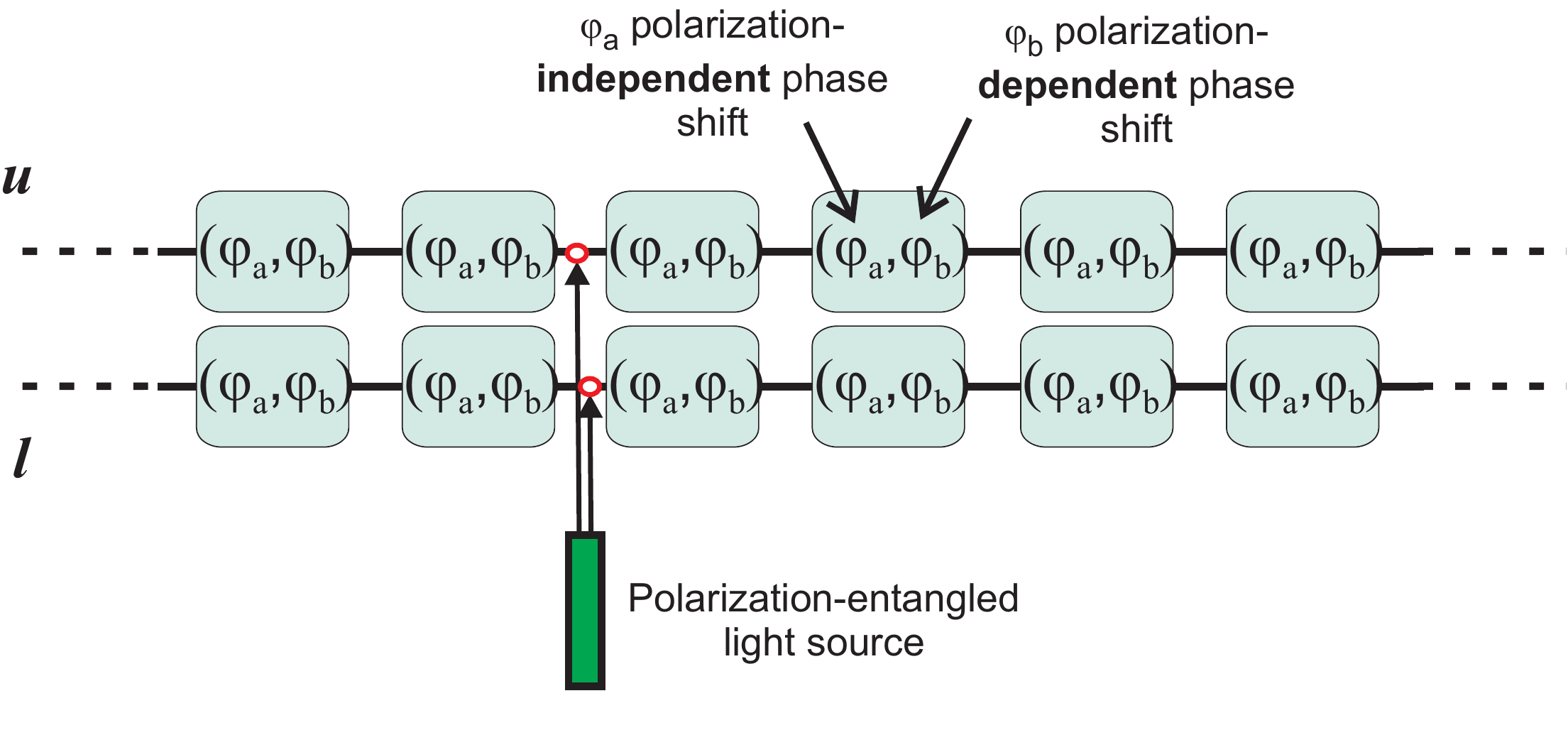}
\caption{Producing winding-number-entangled two-photon states. Each unit cell consists of two diamond graph units,
and so contains a total of four three-ports and two phase shifters. The red circles are units containing an optical switch and an
optical circulator \cite{sim2}; these are used to couple photons in and out of the system and are switched off during operation.}
\label{dualchainfig}
\end{figure}

Consider two chains of unit cells like those of Fig. \ref{schemefig}, distinguishing the upper ($u$) and lower ($l$) chains (Fig. \ref{dualchainfig}). Using the
states $|\Psi^\pm\rangle$ as input, each down conversion photon is directed into one of the two chains, so that the labels $1$ and $2$ in Eq. \ref{bell1} now
correspond to $u$ and $l$. The photons may be coupled into the system via a set of optical circulators and switches, as described in \cite{sim2}. The circulators
are used only to couple photons into the system initially, and to couple them out for measurement at the end; they play no role in the actual operation of the
system in between. We take $\phi_b$ to be polarization-dependent, but we may assume that the action of the phase shifters producing $\phi_a$ are independent of
polarization. In this way, it is arranged for $H$ states to encounter equal phases $\phi_a=\phi_b$, while $V$ states see $\phi_a\ne \phi_b$. The
polarization-dependent phase shifts are easily implemented with thin slices of birefringent material or, if real-time control of the phase shifts is desired,
with Pockels cells.	In the visible and near infrared, it is easy to find crystals with low absorption and strong birefringence; calcite is one example. Other materials with similar properties can be found for other spectral ranges. So it should be relatively easy to produce the necessary phase shifts with negligible effect on performance. The use of electro-optical effects can enable fine adjustment of the phase shift in each cell if necessary.

 If $\phi_b$ is chosen correctly, then the $H$ part is put into a state with winding number $\nu =0$ and the $V$ part into a state with $\nu
=1$. Then the vertically- and horizontally-polarized states will be eigenstates of Hamiltonians with respective winding numbers $\nu_V=1$ and $\nu_H=0$. The final state is therefore
of the form
\begin{equation}{1\over\sqrt{2}}\left( |0_H\rangle_u |1_V\rangle_l \pm |1_V\rangle_u
|0_H\rangle_l\right),\label{entbulk}\end{equation} where the $0$ and $1$ represent winding number values of the Hamiltonians that govern their time evolution, while $u$ and $l$
denote the spatial modes in the upper and lower chains. The state is written in condensed form here; a more explicit expression, including the spatial dependence
of the wavefunction is given in the appendix. The photons now form winding number-entangled qubit pairs.
More generally, both $\phi_a$ and $\phi_b$ may both be allowed to be polarization
dependent; all that matters is that the polarization-dependent combination
$(\phi_a,\phi_b)$ leads to each polarization state seeing a Hamiltonian of different winding
number.

Usually, the global property of winding number is difficult to determine by local measurements. This is especially true for single-photon states which are usually
destroyed by the measurement process, so that multiple measurements cannot be carried out to determine the global state. But here polarization and winding number
remain coupled. This jointly-entangled structure allows the variables to play disparate roles: the discrete winding number leads to topologically-enforced
stability, while polarization, being defined locally, makes the photon state easy to measure. Polarization can be determined by a single local measurement,
allowing the global winding number to be inferred from its value.

Suppose that a perturbation occurs to the system. Normally, this might cause the photon's polarization to change (a polarization-flip error). For example, a
horizontally-polarized state of winding number $\nu$ might try to flip into a vertical state: $|\psi_\nu\rangle_H \to |\psi_{\nu^\prime}\rangle_V$, where
$\nu^\prime$ is the final winding number. But as discussed in more detail in \cite{simwave} and in the appendix, unless the perturbation is strong enough to
severely and globally alter the entire system, transitions from eigenstates of one bulk Hamiltonian to those of a Hamiltonian of different winding number are
suppressed. If the hopping parameters are chosen appropriately, the amplitudes for these transitions can be made arbitrarily small. This means that, to a high
level of certainty, the initial and final winding numbers can be assumed to be equal: $\nu^\prime =\nu$. However, due to the way the system was constructed,
there are no allowed states that have $V$ polarization and which propagate according to a Hamiltonian of winding number $\nu$, so the polarization flip is suppressed.

Thus, barring extreme alterations of the system, polarization-flip errors are greatly reduced.  The suppression of polarization errors occurs without loss of
photons, and so there is no damage to
any coherence or entanglement present in the system. 


\section{Topologically-protected quantum memory registers}

A basic ingredient needed for quantum computing is a quantum memory unit capable of storing a logical qubit $\alpha |0\rangle +\beta |1\rangle$.
Such units need to have read/write capability and should be resistant to bit-flip
errors. This can be achieved by a variation on the strategy above. Once again, topological stability suppresses errors, with polarization used for reading and
writing stored values.
%

A schematic depiction of the memory register is shown in Fig. \ref{ring1fig}.
As before, assume that $\phi_b=\phi_a$ for horizontal polarization and
$\phi_b\ne \phi_a$ for vertical. When a horizontally-polarized photon enters
the ring it is associated with winding number $\nu =0$, but for appropriate values of
$\phi_b$ a vertically polarized photon will have $\nu =1$. The winding number
then serves as the logical bit being stored. Readout of bit values requires
only simple polarization measurements. Since the input photon may be in any
arbitrary superposition of polarization  states, the ring can be used to
store any possible qubit value. In general, an input polarization qubit
$\alpha |H\rangle +\beta |V\rangle$ is stored in a
winding-number/polarization qubit, $\alpha |0_H\rangle +\beta |1_V\rangle$,
where $0$ and $1$ are winding number.

Since photons at normal energies do not mutually interact to a significant degree, multiplexing is possible. Multiple qubits can be stored on a single ring by
using photons of different frequency; addressing the desired qubit then simply requires opening an exit channel from the ring and using a filter or dichroic
mirror with the appropriate frequency-transmission range. Reading out a qubit value consists of measuring the polarization.

\section{Entangled quantum memory registers}

\begin{figure}
\centering
\includegraphics[totalheight=2.0in]{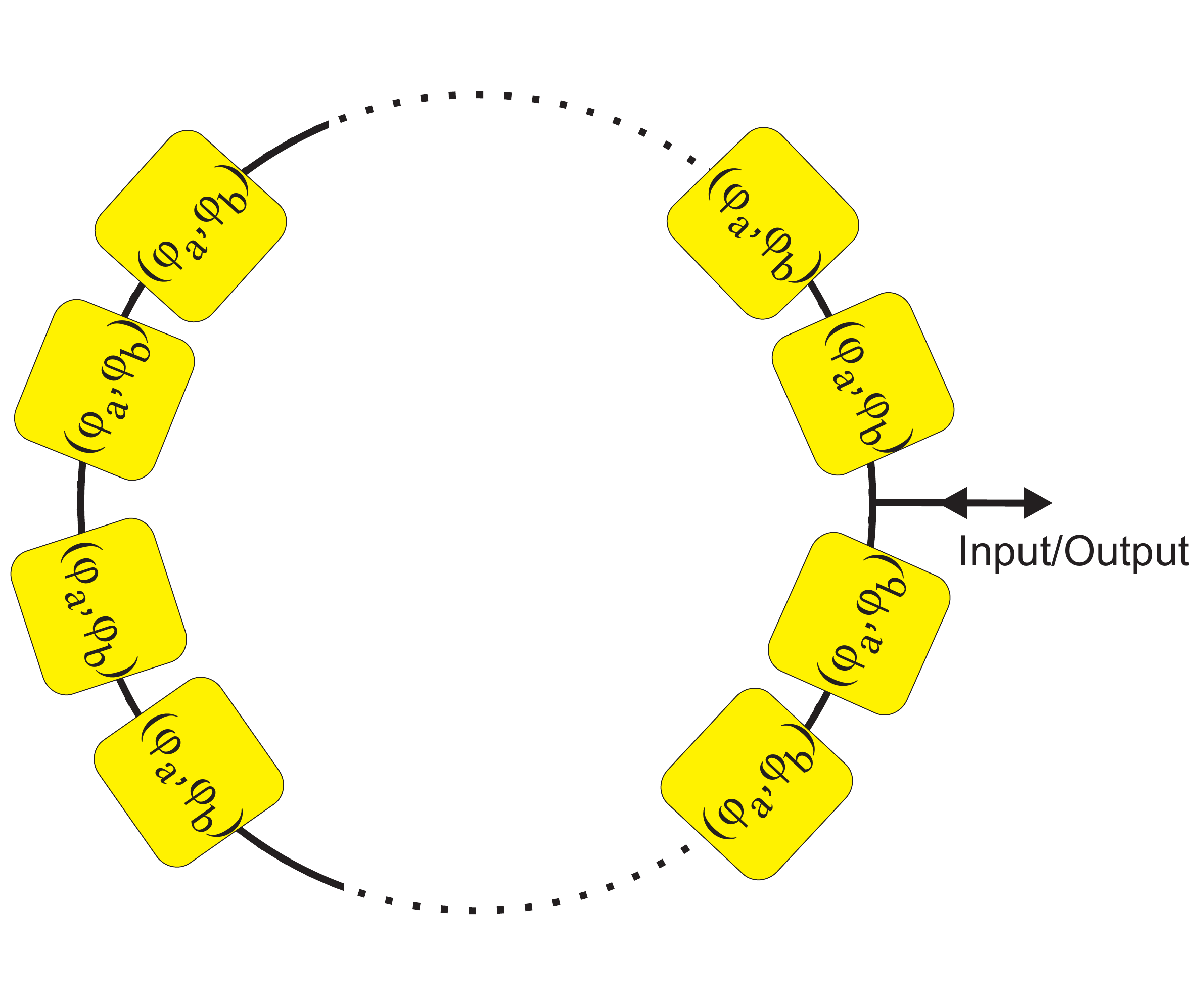}
\caption{Schematic diagram of a quantum memory register. $\phi_b$ is
polarization-dependent, while $\phi_a$ is not. Logical bits are stored in the
winding number of the state and retrieved via polarization measurement.}
\label{ring1fig}
\end{figure}

Notice that the register of Fig. \ref{ring1fig} consists of one strand of the
structure of Fig. \ref{dualchainfig} wrapped into a circle; the
compactification to a circle makes its use in a larger system more practical,
but is not necessary for operation. Each strand of Fig. \ref{dualchainfig} is
already capable of serving as a quantum memory. One could use \emph{both}
strands of Fig. \ref{dualchainfig} (either in the original linear
configuration or compactified to a double-circle structure); for the
polarization entangled input of Eq. \ref{bell1} the result would be two
entangled quantum memory registers, with error suppression provided by the
winding number entanglement.

Such an entangled quantum register could provide novel applications in quantum computing. As one example, using an entangled pair of memory registers would enhance security; by using a subset of the registers for security checks instead of for computation, a data breach would be detectable as a sudden loss of entanglement. Similarly, if a register malfunctions then the location of the malfunction should be easy to track down through the drop in the degree of entanglement of the corresponding pair.

%

\section{Topologically-protected entangled boundary states}\label{topbound}

\begin{figure}
\centering
\includegraphics[totalheight=1.6in]{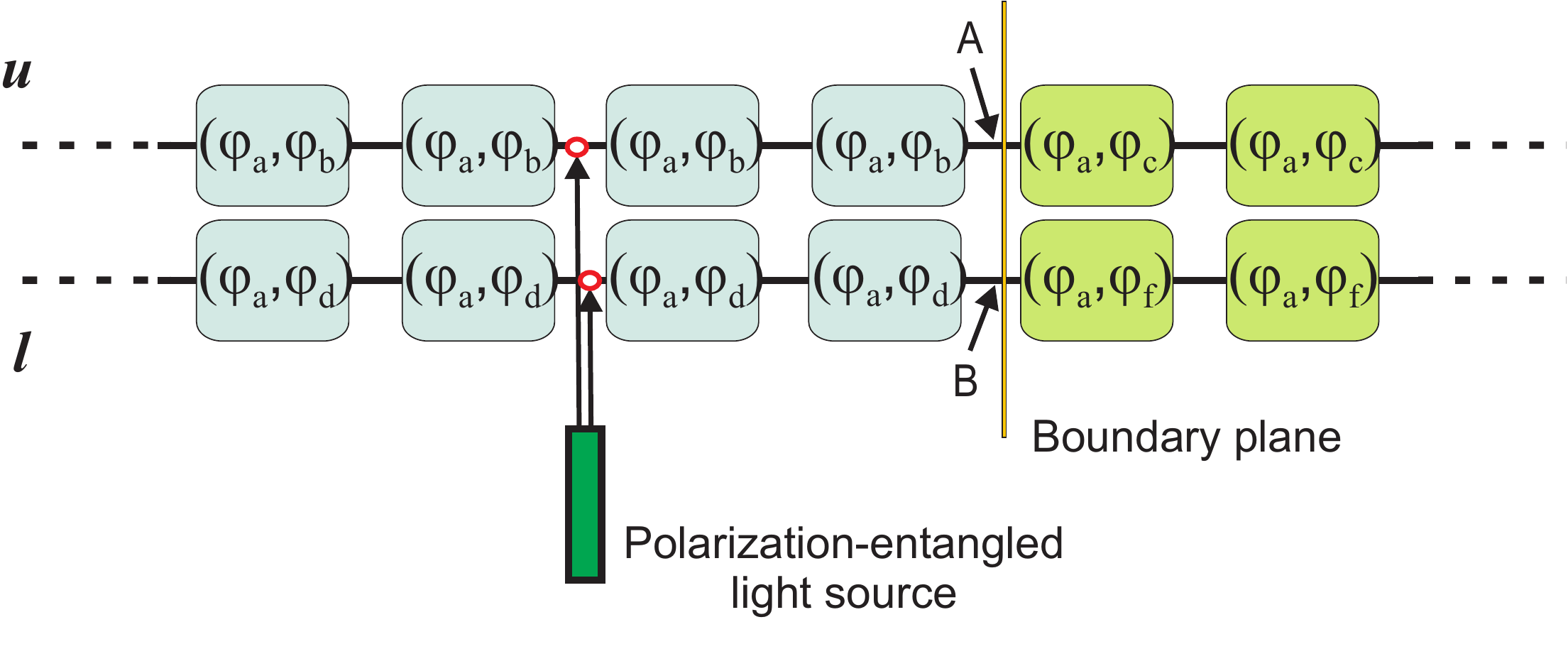}
\caption{Entangled edge states. The points $A$ and $B$ form a boundary between
bulk regions of different winding number. Polarization-entangled input states
lead to winding-number entangled states at $A$ and $B$. }
\label{entedgefig}
\end{figure}

The setup of Fig. \ref{dualchainfig} can be generalized to produce one further effect. First, the polarization-dependent phase can be made different in the upper
and lower branches ($\phi_b$ in upper branch and $\phi_d$ in lower). Then, a boundary plane can be introduced perpendicular to the chains, such that the
polarization-dependent phase will change suddenly when the plane is crossed ($\phi_b\to \phi_c$ in upper branch and $\phi_d\to \phi_f$ in lower), as shown in
Fig. \ref{entedgefig}. As shown in \cite{sim3}, if the phase values on the two sides of the boundary are chosen correctly, then topologically-protected states
appear that are tightly localized on the boundary point. Unlike the approximate winding number preservation in the bulk, the topological protection of the
boundary state is exact and has been demonstrated in  a number of different solid state and optical systems.

Taking the simplest case, suppose $\phi_b=\phi_d$ and $\phi_c=\phi_f$, so that the upper and lower chains are identical. Then for polarization-entangled input
(Eq. \ref{bell1}), the boundary state will be in a superposition of two positions (points $A$ and $B$), as a Schrodinger cat state. These entangled boundary
states will be vertically-polarized and would be topologically protected. Considering just the state at the boundary, let $|e\rangle$ and $|\varnothing \rangle$
represent, respectively, the presence or absence of a localized edge state. Then the state on the boundary plane will be of the form
\begin{equation}{1\over \sqrt{2}}\left( |e_V\rangle_u |\varnothing_H\rangle_l \pm |\varnothing_H\rangle_l |e_V\rangle_u  \right) ,\label{edge1}\end{equation}
where we assume as before that vertical polarizations see different winding numbers on the two sides of the boundary and horizontal polarizations do not. Here,
as before, $u$ and $l$ label whether the spatial mode is in the upper and lower branch. Note that this entanglement is distinct from path entanglement; photons
exist simultaneously in \emph{both} branches, even if edge states are absent from a given branch.

Another possibility is to take $\phi_a\ne \phi_b$ for the \emph{vertical} polarization in the upper chain, but in contrast to take $\phi_a\ne \phi_b$ for
\emph{horizontal} polarization in the lower chain; in this case, there would be an entangled state which is a superposition of having either localized boundary
states at \emph{both} $A$ and $B$ or at \emph{neither}:
\begin{equation}{1\over \sqrt{2}}\left( |e_V\rangle_u |e_H\rangle_l \pm |\varnothing_H\rangle_u |\varnothing_V\rangle_l \right) .\label{edge2}\end{equation}
The states of Eqs. \ref{edge1} and \ref{edge2} are maximally entangled, with entropy of entanglement equal to 1, and may be thought of as topologically-stable implementations of Bell states; these states
can also be used to store entangled qubits.

Edge states appear due to interference between various amplitudes for quantum walks through each chain; they should survive as long as the photons remain
contained inside the system, coherent and unmeasured. Small perturbations in the refractive index or in path lengths along the photon trajectories should not disturb the results. For example, in the appendix numerical results are displayed (Fig. \ref{resfig}) that show that the boundary state persists over a wide range of $v$ and $w$ parameters, as long as the winding numbers do not change.

%
%

\section{Conclusion}

We have proposed a hybrid strategy for quantum information processing, in which local and global properties (polarization and winding number) are
jointly-entangled, allowing one to simultaneously exploit the benefits of both: discreteness of global, topological properties affords stability and error
suppression, while local properties are easy to manipulate and measure. This has applications in producing entangled topological states and in designing quantum
registers (possibly in entangled pairs) with topologically-assisted reduction of bit-flip errors.

Besides reducing bit flip errors, the use of discrete topological quantities also helps maintain loss of entanglement through the same mechanism: if there are no
non-entangled joint states that a photon pair can scatter into, then the entanglement will remain robust. This can help avoid some of the problems that occur in
many approaches to quantum computing as a result of the fragility of entangled states.

Efficient measurement of topological quantum numbers has been a longstanding problem. Although other methods of measuring topological variables in photonic
systems have been proposed or carried out \cite{longhi,ozawa,hafezi,taras,bark,maffei}, they require determination of probability amplitudes by measurements on multiple
photons. The method given here has the advantage of being able to operate at the single photon level.

\ack This research was supported by the National Science Foundation EFRI-ACQUIRE Grant No. ECCS-1640968, AFOSR Grant FA9550-18-1-0056, and by the Northrop Grumman NG Next.

\appendix
\section{Winding number, wavefunctions, and topological state protection}

\begin{figure}
\begin{center}
\subfigure[]{
\includegraphics[scale=.4]{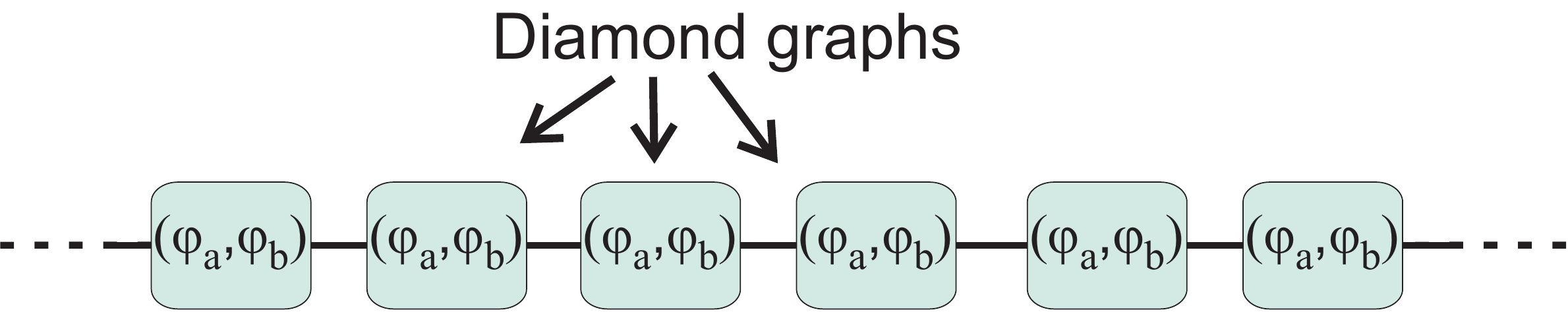}}
\subfigure[]{
\includegraphics[scale=.4]{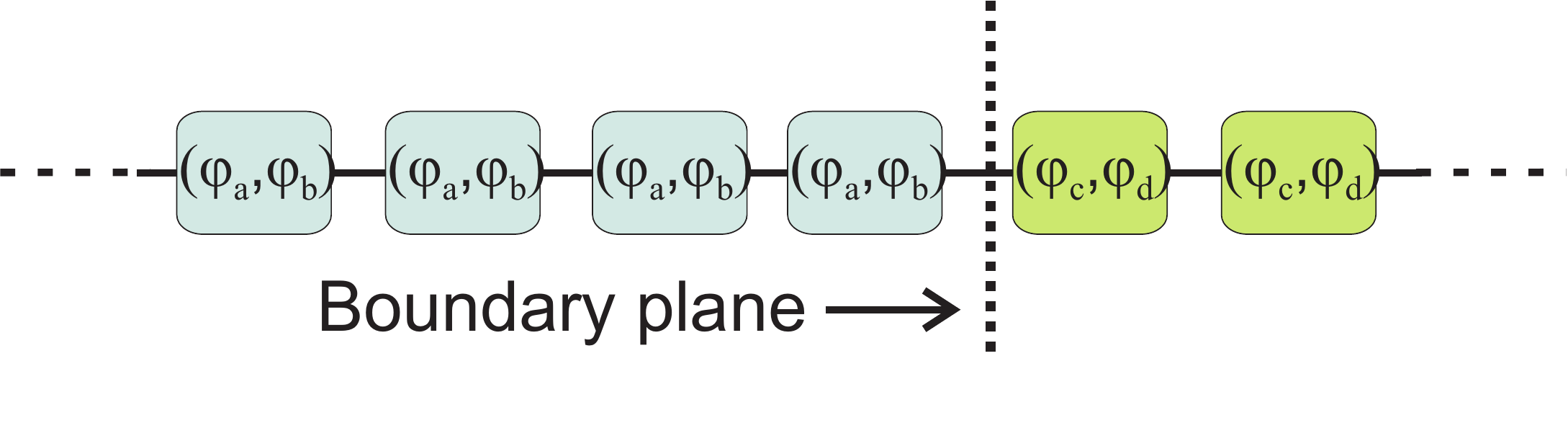}}
\caption{(a) A linear periodic lattice formed by diamond graphs,
with alternating phases $\phi_a$ and $\phi_b$. The phases control the transition
amplitudes $w$ between lattice sites (the rectangular boxes) and $v$ between
subsites $A$ and $B$ (the two diamond graphs within each lattice site) vary.
$A$ and $B$ play the role of substates at each site. (b) Two such chains connected
end to end. For some values of phase shifts, the two chains will support states of
different winding number, with stable localized states appearing at the boundary.
}\label{chainfig}
\end{center}
\end{figure}

In Ref. \cite{sim3} it was shown that the chain of directionally-unbiased multiports in Fig. \ref{chainfig}(a) is a photonic equivalent of the
Su-Schrieffer-Heeger (SSH) system used to model the behavior of polymers. In this appendix, we briefly review the topological properties of this system and
verify via numerical simulations that these properties hold for chains of unbiased multiports; in particular, we numerically demonstrate the resistance of the
wavefunction to enter regions of different winding number.

The one-dimensional SSH system is composed of a periodic string of unit
cells, with two alternating subcells $A$ and $B$ of distinct reflection and
transmission amplitudes within each cell. This forms a two-level system, with
a momentum-space Hamiltonian of the form
\begin{equation}\hat H= \int_{-\pi}^\pi dk\; E(k)\; {{\bm d(k)\cdot \sigma}\over {|\bm d(k)|}} \otimes
|k\rangle \langle k| ,
\end{equation} where $k$ is the quasi-momentum and the integral is over a
full Brillouin zone.  Due to the existence of a chiral sublattice symmetry
with generator $\Gamma = e^{-i{\pi\over 2}  \sigma_z}$ that anticommutes with
the Hamiltonian, the spectrum is symmetric, with energies coming in
opposite-sign pairs, $\pm E(k)$. The gap between the two energies only closes
when $A$ and $B$ have equal transmission amplitudes.

The Hamiltonian is determined by the vector \begin{equation}\bm
d(k)=\left(v+w \cos k\right) \hat x + \left(w\sin k\right) \hat y
 ,\end{equation} where $v(k)$ and $w(k)$ are respectively the intracell hopping amplitude between $A$ and $B$
and the intercell hopping amplitude. $\hat x$ and $\hat y$ are basis vectors in the two-dimensional Hilbert space spanned by the $A$ and $B$ substates. In the
simplest SSH model, $v$ and $w$ are constants, and $d\bm (k)$ traces out a circle in $k$ space; for the unbiased multiport chains, $v$ and $w$ are continuous
functions of $k$, so that the circular paths become continuously deformed. Since topological properties of systems are unchanged by continuous deformations of
the parameters, the unbiased multiport system has the same topological properties as the SSH model, as is verified numerically below.

$\bm d(k)$ must remain orthogonal to the chiral symmetry generator $\sigma_z$ in order to preserve the symmetry. However, the direction of the vector $\bm d(k)$
becomes undefined at $k=\pi$ when $v=w$. The gap between the energy levels closes at the parameter values for which this occurs. For other parameter values, $\bm
d(k)$ must avoid the origin, leading to a distinction between values at which the path traced out by $\bm d(k)$ encloses the origin and those for which it does
not. The latter cases are topologically trivial, with bulk winding number $\nu=0$, while the former cases have nontrivial winding number $\nu =1$. The winding
number cannot change without $\bm d(k)$ crossing the singular point and the energy gap closing.


When the Hamiltonian changes abruptly from one topological state to different one (say from winding number $0$ to winding number $1$), highly-localized states
appear at the boundary between the two topological regions. It has been demonstrated in a number of different physical contexts that there is a form of
topological protection attached to these states: no continuous localized disturbance can destroy the state or cause a change in the winding number on the two
sides of it. In particular, this has been demonstrated experimentally for a number of photonic systems \cite{lu1,lu2}, including systems based on photonic
quantum walks \cite{kitrev,broome}.

Further, it can be shown \cite{simwave} that a wavefunction initially present in the bulk on one side of the boundary tends to resist transmission into
the second, topologically distinct, bulk region. The wavefunction instead shows a tendency to reflect back into the original region when it encounters the boundary. This tendency can be made nearly complete by a wise choice of the parameters in the Hamiltonian, as will be discussed below in the context of the SSH system.

This topologically-assisted suppression of transitions is the key to why the systems in Sections \ref{hyperbulk}-\ref{topbound} are of interest. The polarization
will be linked with winding number, so that the suppression of transitions between different winding number states will suppress polarization changes.



Label each unit cell of the lattice by an integer position label $n$. Each such unit cell has two subunits or ``substates'', labeled $A$ and $B$. We take the
coordinate system such that the center of each cell is at $x=n$, with the $A$ and $B$ subcells located at $x=n-{1\over 4} $ and $x=n+{1\over 4}$, respectively.
We insert a photon at some initial site and then let it undergo a quantum walk. We assume the insertion point is at an initial $A$ subsite; corresponding
expressions for insertion at $B$ are similar. The Hamiltonian can be expressed \cite{simwave} in the form
\begin{equation}H(k)=E_k\left(
\begin{array}{cc}0 & e^{i\theta_k-ik/2} \\ e^{-i\theta_k+ik/2} & 0\end{array}\right) ,\label{hampolar}
\end{equation}
where \begin{equation}E_k = \left( v^2+w^2 +2vw\cos k \right)^{1/2} \end{equation} and \begin{equation}\theta_k =  \tan^{-1} \left({{(v-w)}\over {(v+w)}}\tan
{k\over 2}\right) .\label{tantheta}\end{equation} The form of Eq. \ref{hampolar} shows clearly the winding of the matrix elements of $H$ in the complex plane as the angle $\theta_k$,
which is essentially a Berry phase, changes. The eigenvectors, of energies $E_\pm =\pm E_k$, are of the form \begin{equation}|\pm\rangle = {1\over \sqrt{2}}\left( \begin{array}{c} 1 \\
\pm e^{-i(\theta_k -{k\over 2})}\end{array}\right) ,\label{lpmupm}\end{equation} where the two components represent the amplitudes of being in the $A$ and $B$
states.

Given a state initially localized at subsite $A$ of site $n_0$, the spatial wavefunction at later time $t$ will be of the form \cite{simwave}
\begin{eqnarray} \psi_{v,w}(r,t)&=& {1\over N}\sum_{kn} e^{ik(n-n_0)} \left( \phi \left( r- n+{1\over 4}\right)   \cos (E_kt)\right. \label{psiAA} \\
& & \qquad\qquad \left. +
ie^{-i\theta_k+{i{k\over 2}}} \phi \left( r- n-{1\over 4}\right)   \sin (E_kt)\right) .\nonumber
\end{eqnarray}
The $\phi$ functions are the Wannier functions \cite{wan,kohn,cloiz}, defined as the Fourier transform of the momentum-space Bloch wavefunctions
\begin{equation}\phi_n(r) = \phi(r-n) ={1\over \sqrt{N}}\sum_k e^{-ikn}\psi_k(r).\end{equation} The Wannier functions are tightly localized near the lattice sites
(labeled by integer $n$), are orthonormal, and form a complete basis set for the allowed position-space wavefunctions. $t$ here is some integer multiple of the time between steps of the discrete quantum walk, $t=mT$, with $m=0,1,2,\dots$. The winding number gives the number of times that the phase $\phi_k$ winds around the circle as $k$ traverses a complete Brillouin zone.

The entangled states in the main text of the paper can be written more explicitly in terms of these wavefunctions. For example, the states of Eq. \ref{entbulk}
can be written
\begin{equation}{1\over \sqrt{2}} \left( |\psi_{v_1,w_1},H\rangle_u |\psi_{v_2,w_2},V\rangle_l \pm  |\psi_{v_2,w_2},V\rangle_l |\psi_{v_1,w_1},H\rangle_u \right)
,\end{equation} where $(v_1,w_1)$ are a pair of hopping parameters corresponding to winding number $0$ and $(v_2,w_2)$ correspond to winding number $1$. The
wavefunctions of Eq. \ref{psiAA} correspond to the overlap between the state vectors and position eigenstates: $\psi_{v,w}(r,t)=\langle r|\psi_{v,w}\rangle$.

We now numerically display the existence of some of the properties mentioned above for the case of quantum walks on chains of directionally-unbiased multiports. First consider a
single chain of such multiports (Fig. \ref{chainfig}(a)) arranged into alternating pairs of diamond graphs, as in Section \ref{hyperbulk}. The amplitudes $v$ and
$w$ in this case are functions of the adjustable phases $\phi_A$ and $\phi_B$ in the two diamond graphs making up each unit cell. The Hamiltonian and the
associated vector $\bm d(k)$ can be readily calculated as functions of these phases \cite{sim3}.

\begin{figure}
\begin{center}
\subfigure[]{
\includegraphics[scale=.26]{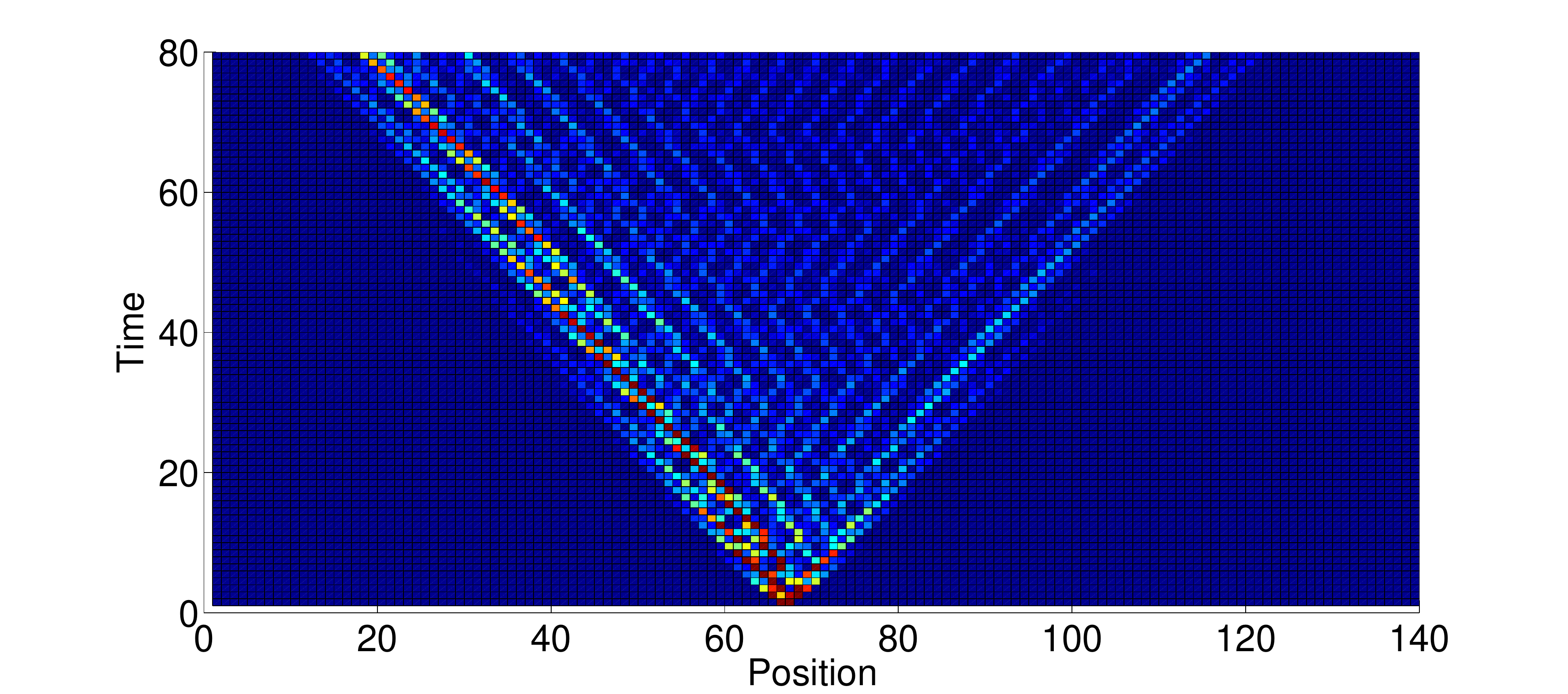}}
\qquad  \subfigure[]{
\includegraphics[scale=.26]{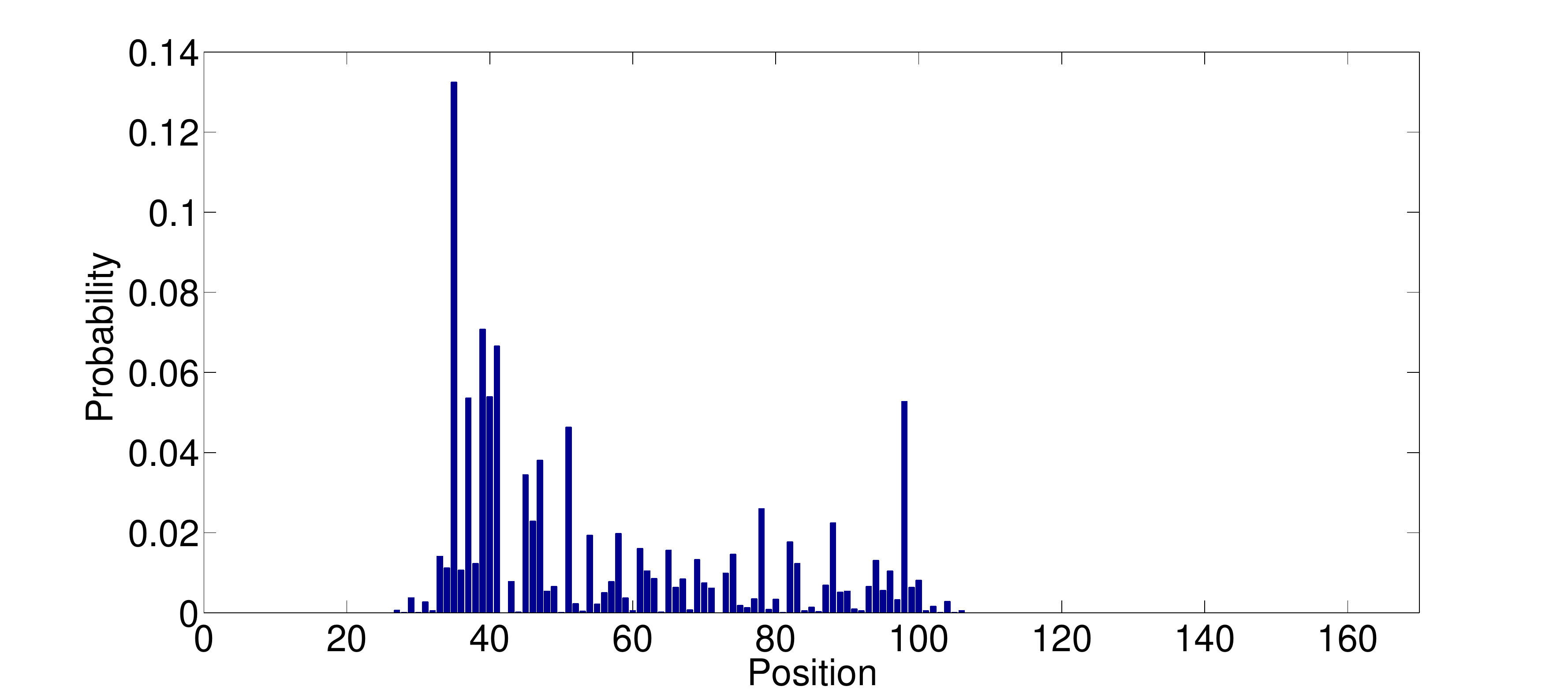}}
\caption{When all of the unit cells of a chain have the same parameter values, a state inserted into the system at any point exhibits standard quantum walk
behavior, with ballistic spreading of probability.
(a) shows the evolution of the spatial probability distribution versus time, while (b) shows the spatial distribution at a fixed time (after 50 time steps).
Here the parameter values used are $\phi_{A} = -\pi/2$ and
$\phi_{B} = 0$, corresponding to a winding number of $1$. The photon starts initially at position 68.
}\label{onechainfig}
\end{center}
\end{figure}

When a photon is inserted into the system at a given location, it will begin
a quantum walk. After a given number of time steps, the probability
distribution can be calculated for the location of the photon. In a classical
random walk, the distribution would be expected to have Gaussian form, with a
width proportional to $\sqrt{N}$, where $N$ is the number of steps. However,
a quantum walk exhibits ballistic behavior, with a distribution that spreads
linearly in time. Calculation of the distribution for the unbiased multiport
chain shows such ballistic behavior, as seen in Fig. \ref{onechainfig}. The bias of the walk (left or right) can be altered by adjusting the values of $v$ and $w$.


\begin{figure}
\begin{center}
\subfigure[]{
\includegraphics[scale=.26]{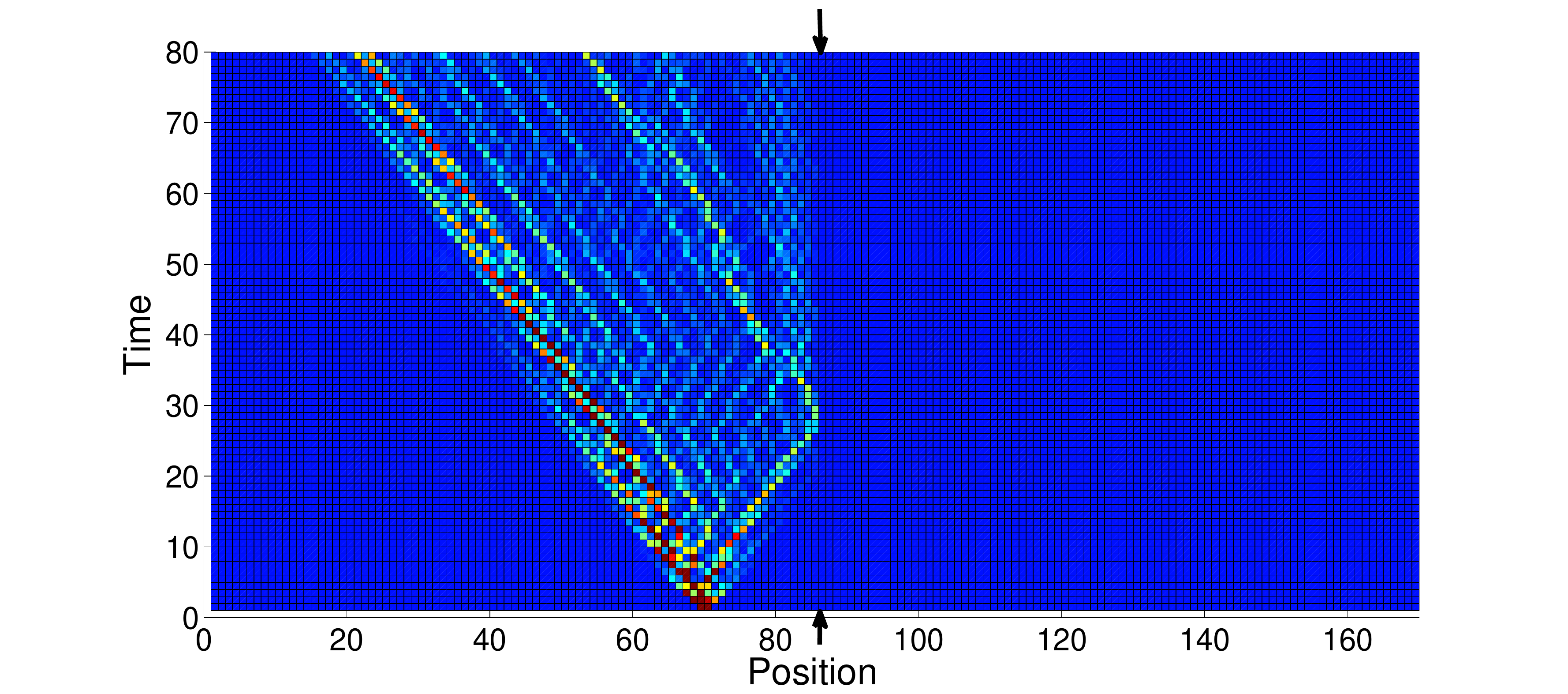}}
\subfigure[]{
\includegraphics[scale=.26]{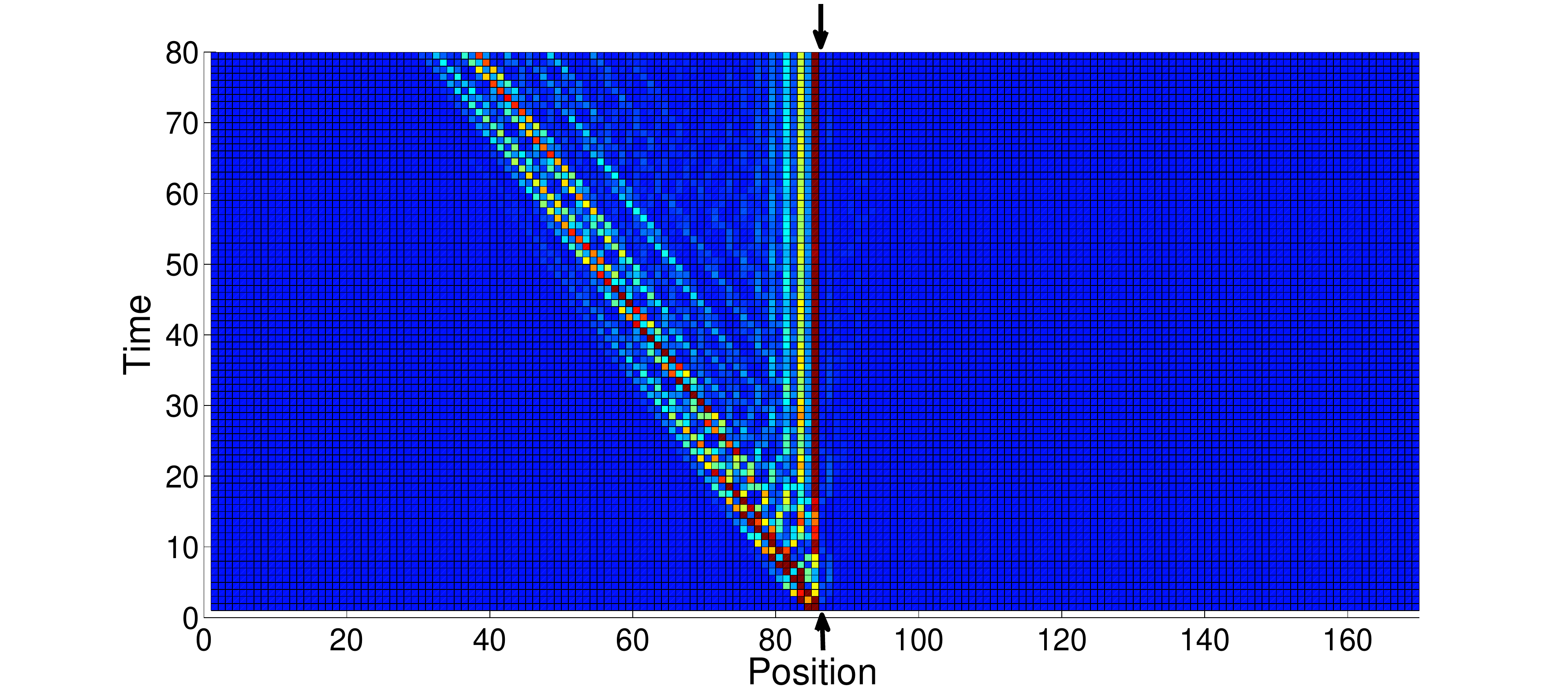}}
\subfigure[]{
\includegraphics[scale=.26]{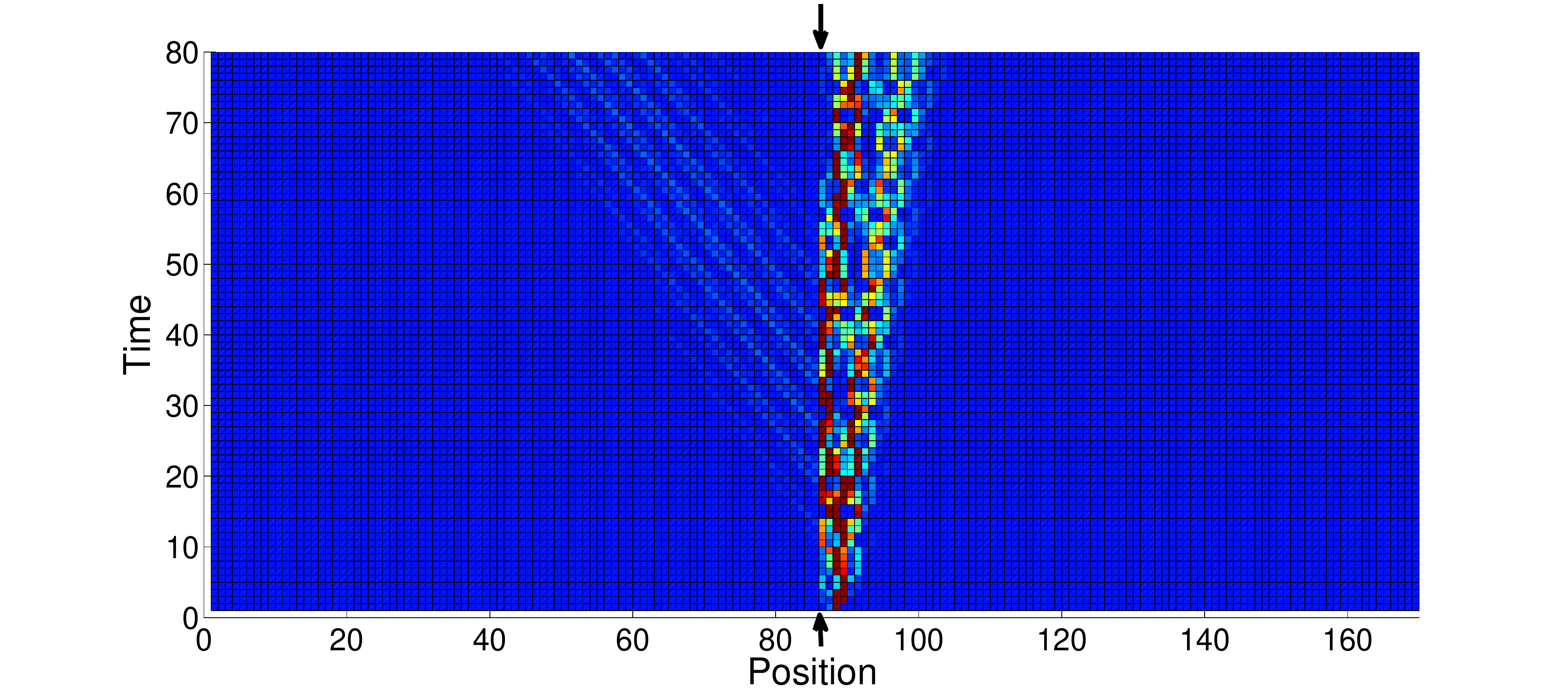}}
\caption{Two unbiased multiport chains with different winding numbers are
connected, with the boundary position 86 (indicated by the arrows). The left-hand chain has parameters
$\phi_{A1} = -\pi/2$,
$\phi_{B1} = 0$
 (winding number $\nu =1$), while the right-hand chain has parameters $\phi_{A2}=1.5$,
$\phi_{B2}=2.5$ (winding
 number $\nu =0$). We see in (a) and (b) that states starting out with winding number
 1 have little amplitude to  cross to the right side of the boundary.
Similarly in (c), winding number 0 states
tend not to cross to the left. We also see that, unlike the topologically trivial
case of the previous Figure, a stable state accumulates at the boundary, as can
be clearly seen in the next Figure. The initial positions in parts (a), (b), and (c) are
respectively 68, 85, and 88.
}\label{twochainfig}
\end{center}
\end{figure}

\begin{figure}
\centering
\includegraphics[totalheight=1.5in]{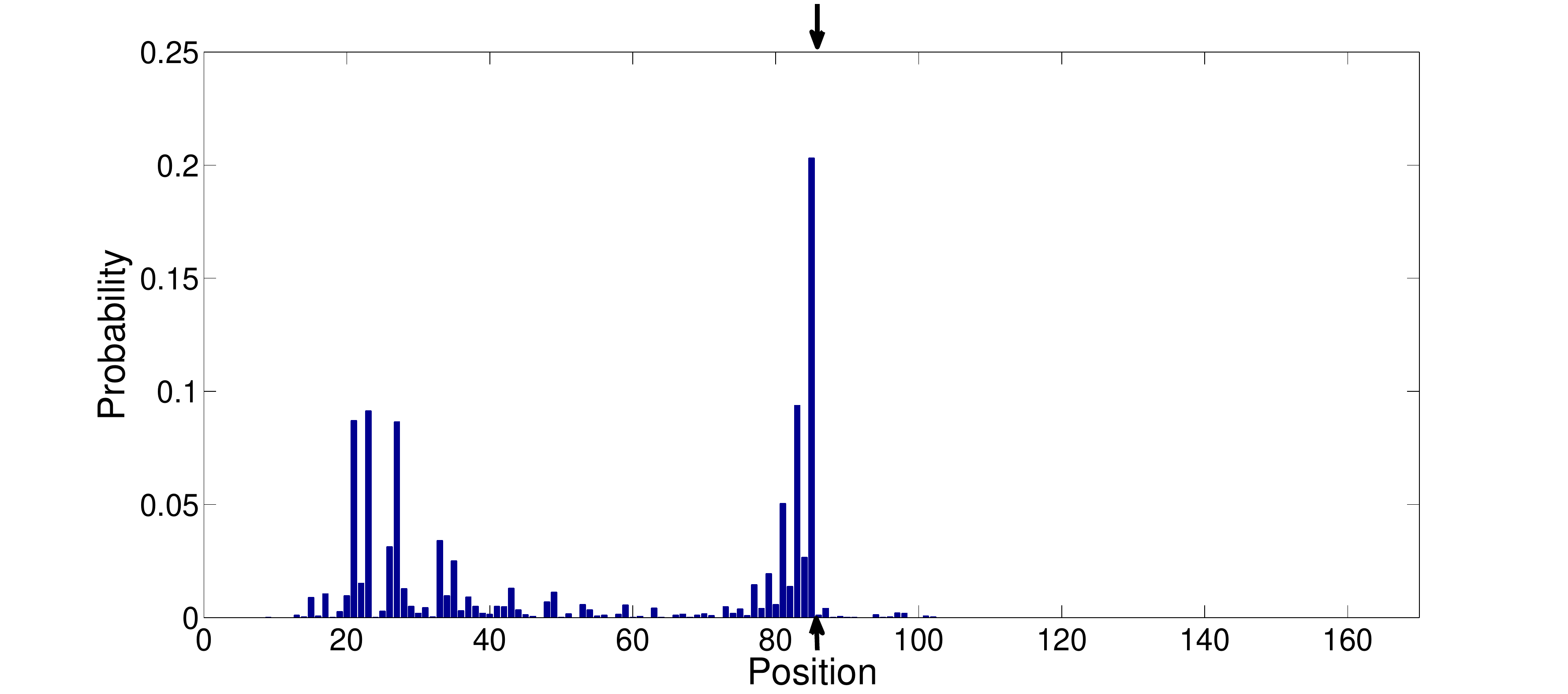}
\caption{A fixed-time plot of the spatial probability
distribution for the distribution (b) of the previous Figure, after $100$ time steps. It is seen
clearly that
the ballistic behavior comes to an abrupt stop as the boundary
between topological phases is encountered (indicated by the dashed line at position 86).
Any amplitude that arrives at the boundary accumulates there. The small amount of amplitude
that crosses the boundary quickly decays to zero.
(Compare to the the boundary-free, topologically homogeneous case in
Fig. \ref{onechainfig}(b), which had the same initial condition.) } \label{twochainhistogramfig}
\end{figure}

Consider now two chains lined up end to end, and connected at their mutual boundary, as in Fig. \ref{chainfig}(b). Adjust the diamond graph phases so that the
winding number is $0$ on the left side of the boundary and $1$ on the right side. As shown in Figs. \ref{twochainfig} and \ref{twochainhistogramfig}, a state
beginning in one region tends not to cross into the other region, even if the available energy levels are the same on both sides. The mismatch of winding numbers leads to a mismatch of eigenstates on the two sides, which in turn
reduces propagation from one side to the other. Fig. \ref{twochainhistogramfig} also clearly shows the accumulation of the localized state at
the boundary, a feature that is absent from the topologically uniform case of Fig. \ref{onechainfig}(b). The peak that accumulates at the boundary remains fixed
at that location for all time.

\begin{figure}
\centering
\includegraphics[totalheight=1.6in]{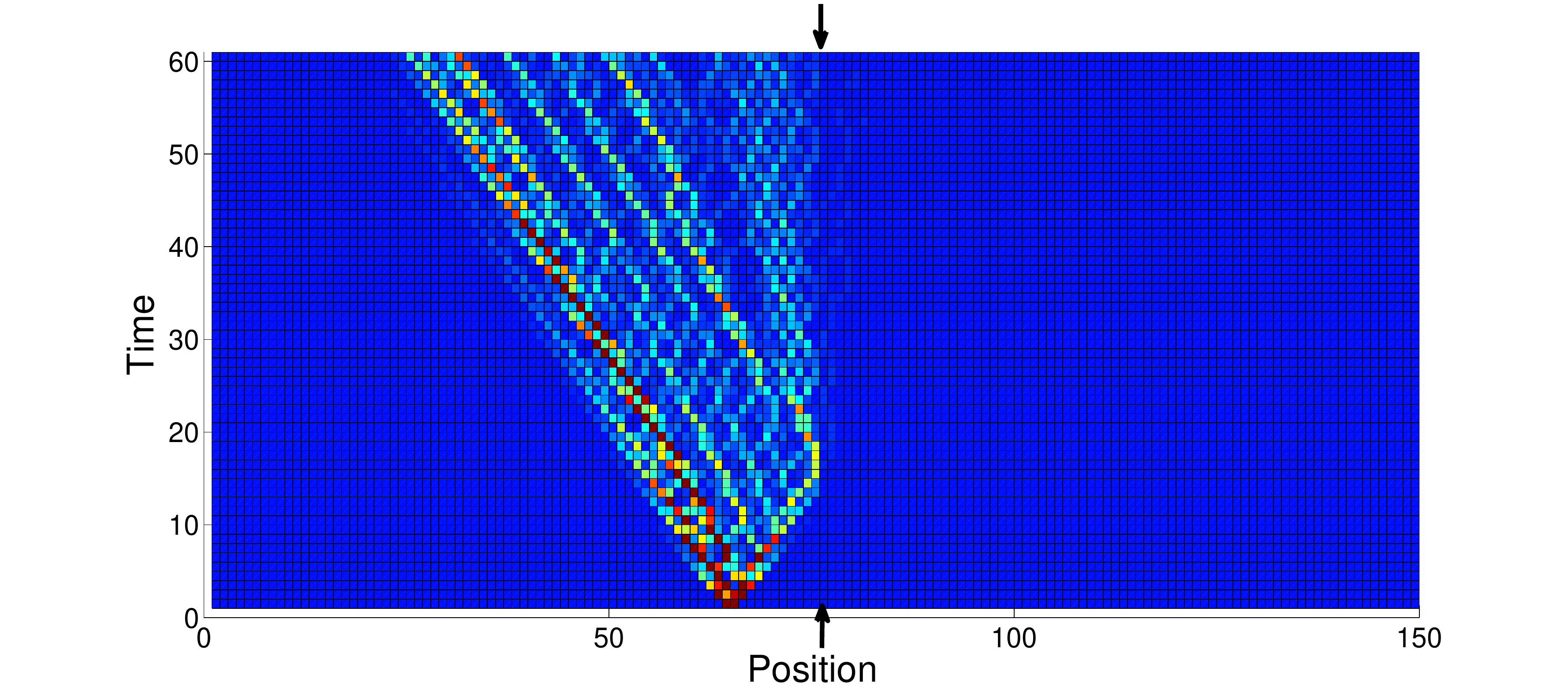}
\caption{Effect of a perturbation to the system. Two chains with
different winding number are again connected, with boundary indicated by the dashed line.
After 30 time steps, the system is given a brief jolt in which the
phase shifts on the left are momentarily altered to the values on the right side
of the boundary, before returning to the original values.  No scattered states
of the wrong winding number propagate away
from the disturbance,
as can be seen by the fact that there is still no amplitude leaking across
the boundary.} \label{disturbfig}
\end{figure}

Unlike the protection of the boundary states, the reduced level of bulk state transitions is partial and depends continuously on the hopping parameters. The
preservation of the bulk state is ``topologically-assisted'' in the sense that it occurs because of reflection at the boundary between topological phases, but it
not of a topological nature in itself, since the transmission and reflection coefficients remain continuous functions of the hopping parameters within each
topological sector. The reflection at the boundary is not total: a small transmission amplitude into the second region exists, but it can be arranged to be
negligible. For the specific case where the two hopping amplitudes $v$ and $w$ interchange the values when the boundary is crossed, $v\leftrightarrow w$, it is
shown analytically in \cite{simwave} that the degree of leakage across the boundary depends on the difference $|v-w|$ between the two hopping amplitudes. The
transmission amplitude is $100\%$ when $v=w$, but drops toward zero as $|v-w|\to 1$. So the leakage into the second region can be made as small as desired by
taking $|v-w|$ sufficiently large. Qualitatively, the principle reason for this is that the change in topology of the Hamiltonian forces a sudden, discontinuous
change in the eigenstates on the two sides of the boundary, making it hard for the rightward-propagating solutions on the two sides to be consistently patched
together: the net result is an increased likelihood of reflection.

Reflection at points where there are sudden changes in the dynamics are a very general occurrence. Not only do they occur at points of sudden potential energy change, as described in every quantum mechanics text, and at points of topological phase change as considered here (see also \cite{duncan} for an alternative approach), but something very similar happens at many other types of sudden inhomogeneities, including boundaries between regions governed by nonrelativistic (Schr\"odinger) and relativistic (Dirac) dynamics \cite{meyer1,meyer2,meyer3}.

Finally, consider a disturbance to the system. Again suppose a two-chain system with different winding numbers on the two chains, but now we perturb the system.
In Fig. \ref{disturbfig}, the result is shown when a state of winding number $\nu =0$ is perturbed after time-step $30$. The disturbance consists of altering the
phase shifts to those characteristic of winding number $\nu =1$ for one time step, then returning to the original values on subsequent steps. The disturbance is
therefore localized in time. Normally, such a sudden jolt to the system would create a scattered state capable of propagating away to infinity in both
directions. But once again, we see negligible propagation into the right-hand region, indicating that no scattered state associated with the Hamiltonian of "wrong"
winding number appears. Other types of disturbances (localized in space, rather than time, for example, or with different values for the perturbed phases) lead
to similar results.

The plots above demonstrate the resilience of the bulk state: the state remains largely unaffected by brief disturbances that temporarily flip the winding number of the Hamiltonian, and tends to reflect to avoid entering regions of opposite winding number. Taken together, this indicates a resistance to states that evolve according to one Hamiltonian making transitions to states evolving according to a Hamiltonian of different winding number; since the Hamiltonians are determined by the polarization states, this indicates by extension a reduced rate of polarization flips.

The resistance to transitions between regions of different winding numbers is quantified in \cite{simwave}, where transmission and reflection coefficients at the
boundary are calculated.  Assuming the simplest case, where the hopping amplitudes are interchanged at the boundary ($v_{left}=w_{right}$ and
$w_{left}=v_{right}$) the transition rate decreases as $|v-w|$ grows. The transition probability per encounter with the boundary can be made to drop below
$10^{-3}$, for example, by choosing the parameters such that $|{{v-w}\over {v+w}}|>.96$.

\begin{figure}
\begin{center}
\subfigure[]{
\includegraphics[scale=.12]{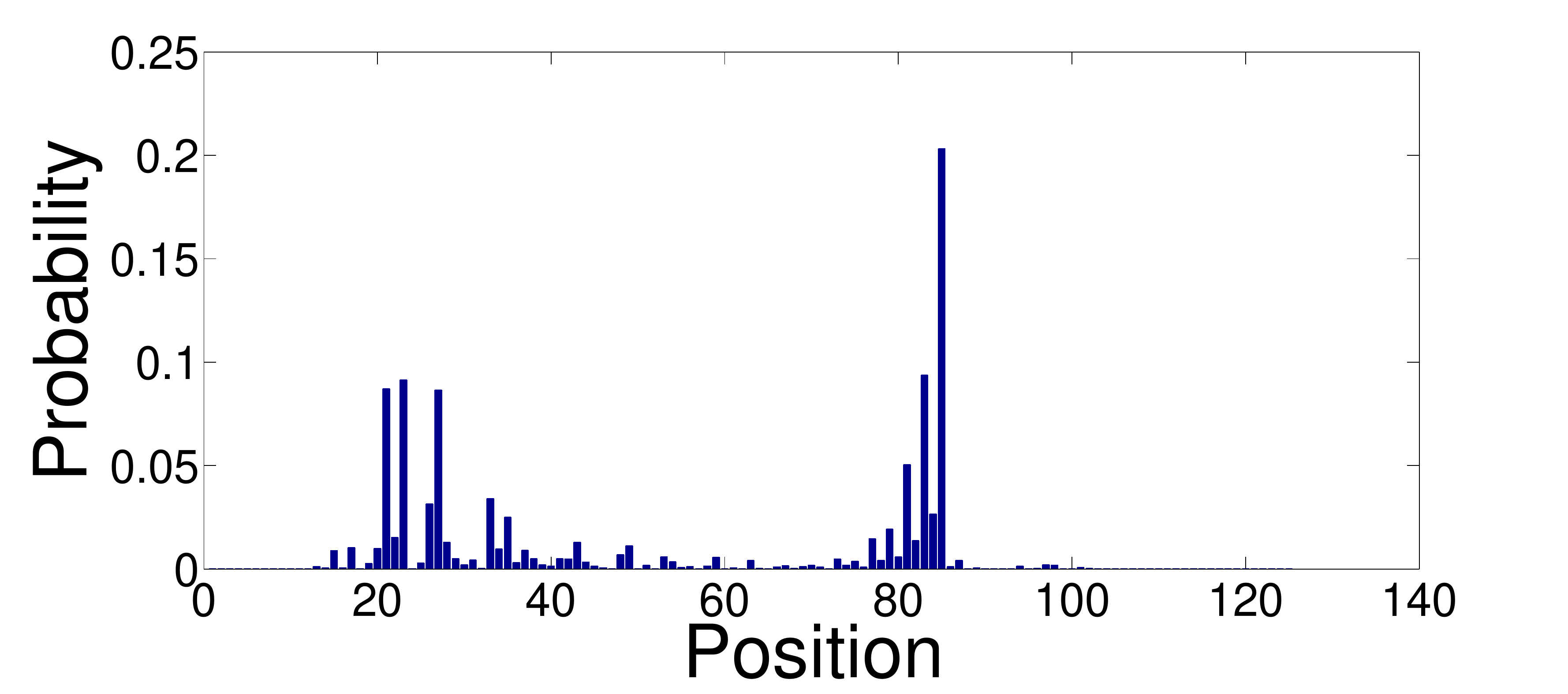}}
\subfigure[]{
\includegraphics[scale=.12]{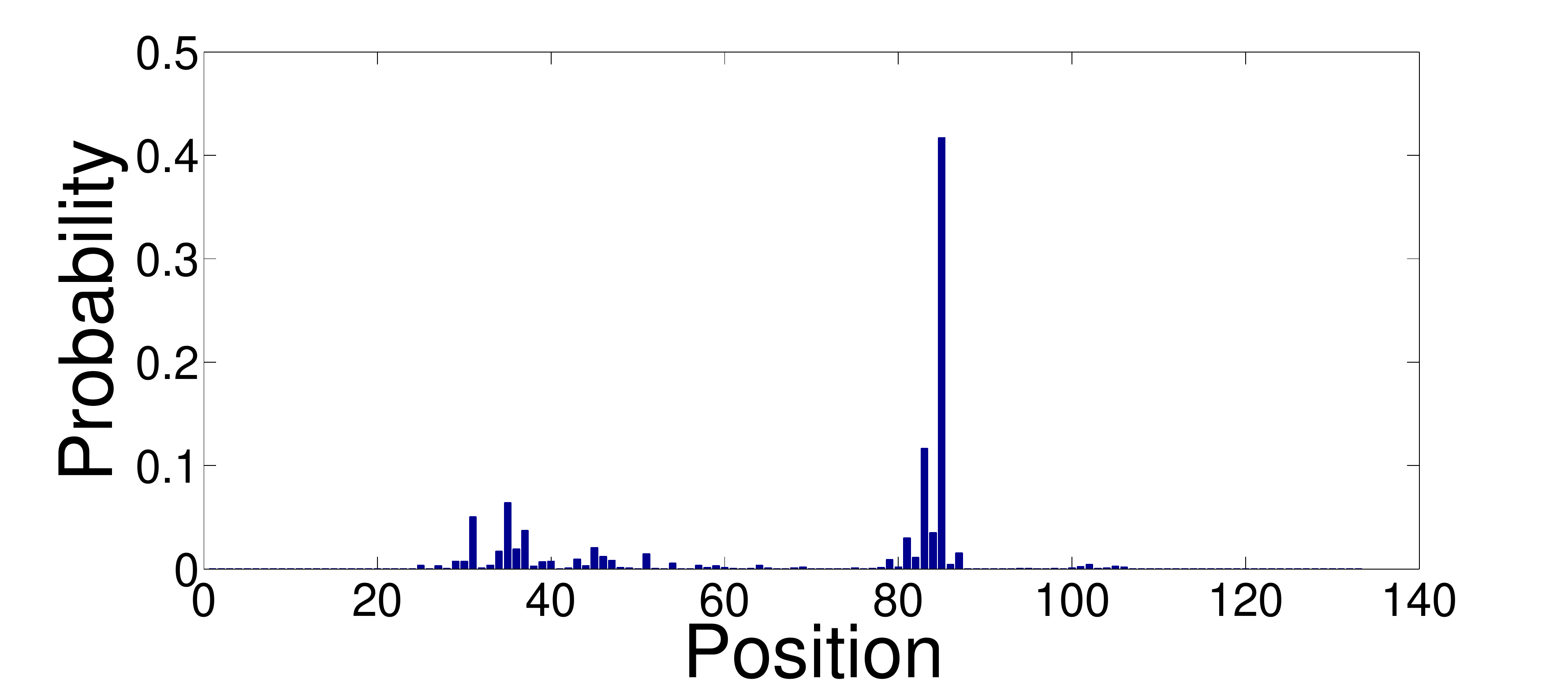}}
\subfigure[]{
\includegraphics[scale=.12]{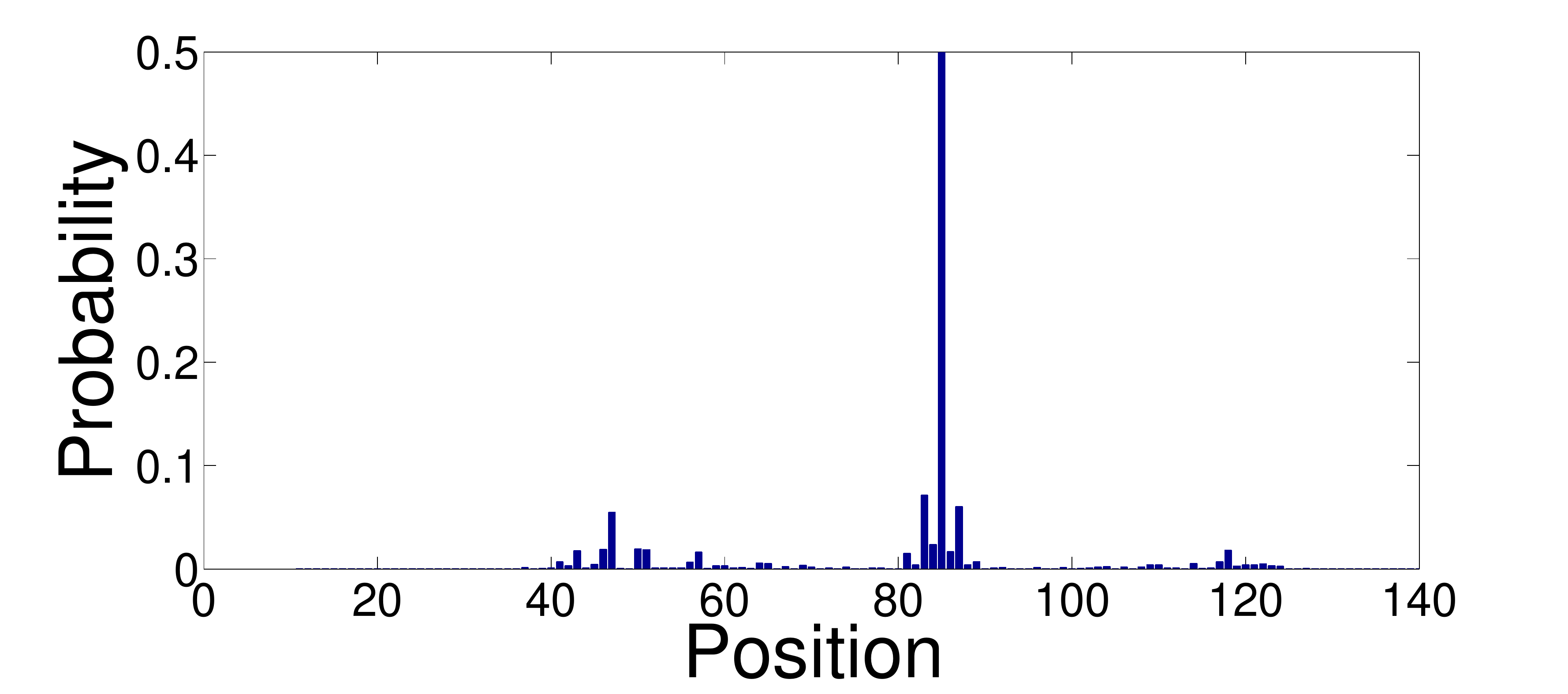}}
\subfigure[]{
\includegraphics[scale=.12]{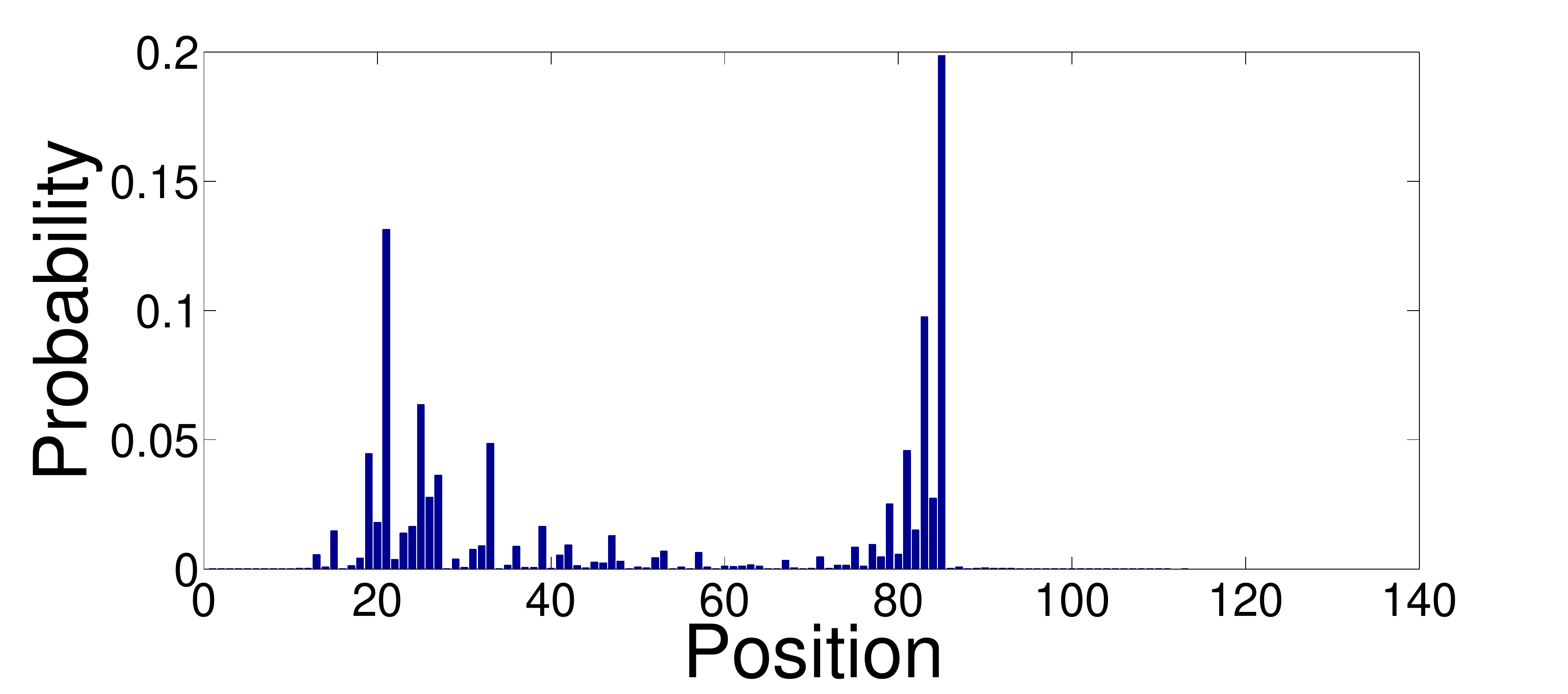}}
\subfigure[]{
\includegraphics[scale=.12]{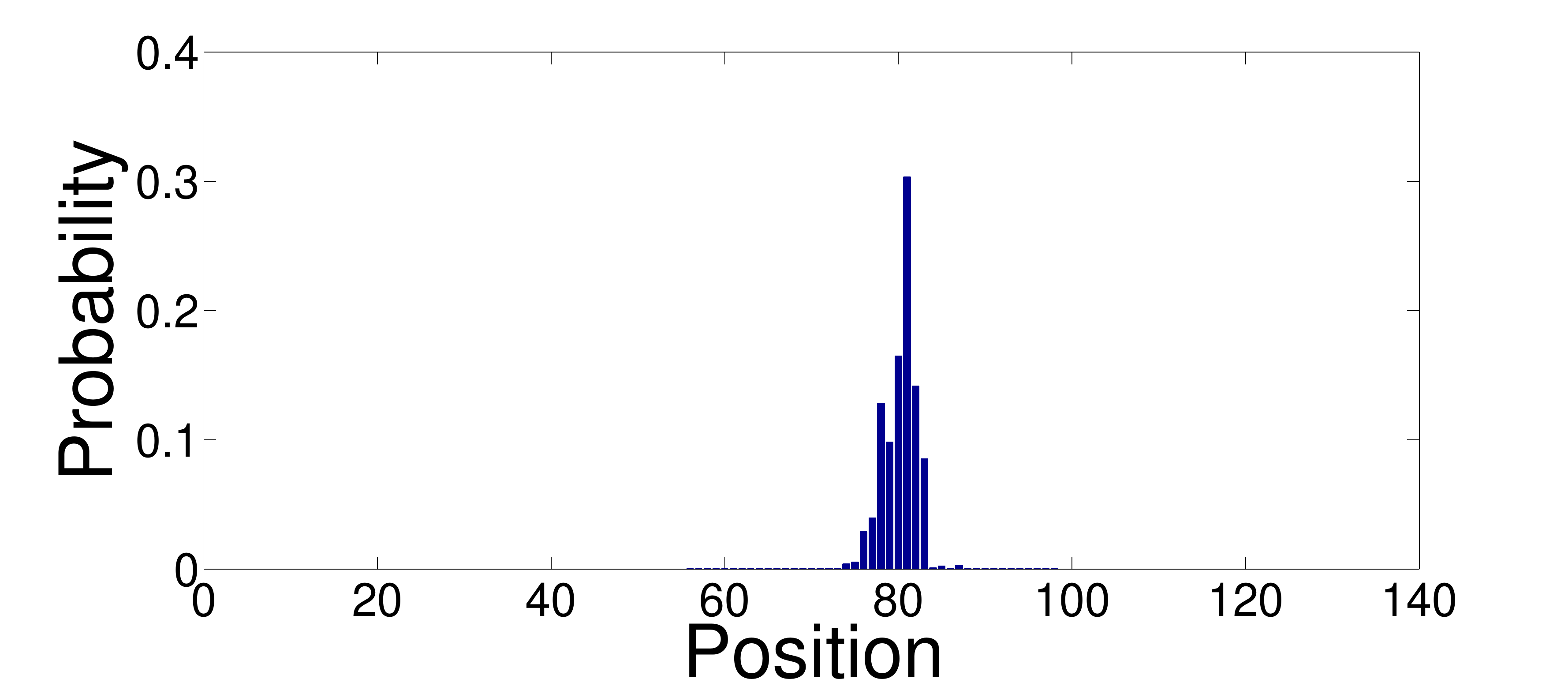}}
\subfigure[]{
\includegraphics[scale=.12]{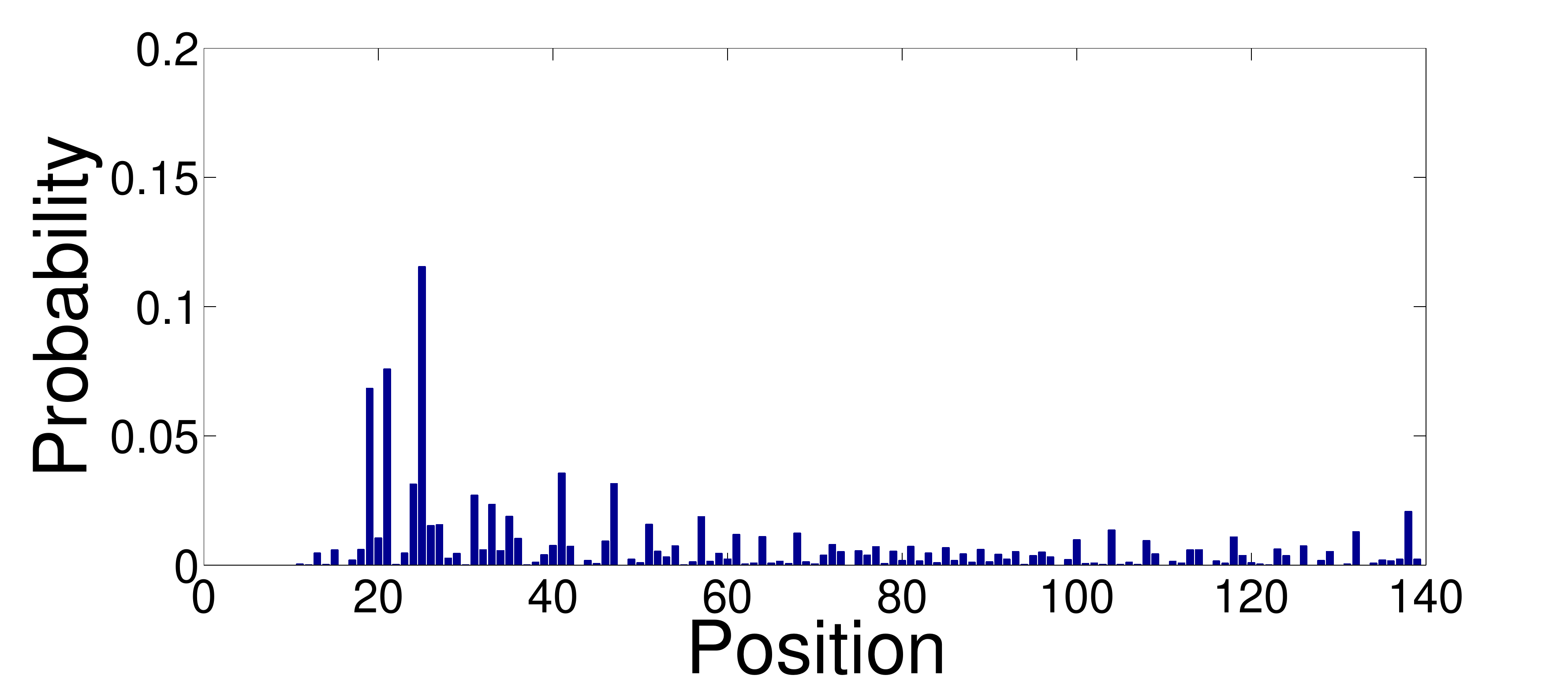}}
\caption{In Figs. (a)-(e), a variety of phase settings are used, all of which lead to different winding numbers on the two sides of site 86. Despite the details of the walk changing with each change in settings, the bound state remains stable and fixed at the boundary. However, when the winding numbers are made equal on the two sides, the bound state abruptly disappears. (All of these figures were computed after 100 time steps.)
}\label{resfig}
\end{center}
\end{figure}

The boundary state, of course, is well known to be stable against perturbations of the Hamiltonian; this can be easily demonstrated. In Figs. \ref{resfig}(a)-(e), a
range of different phase settings are applied. As long as the winding numbers remain different on the two sides, the edge state at the boundary (site 86)
remains. Everything else about the walk dynamics may vary, but existence of the bound state remains highly stable. Only when the singular point of the
Hamiltonian is crossed and the winding number becomes equal on the two sides (Fig. \ref{resfig}(f)) does the boundary state disappear.

%


\section*{References}

\begin{thebibliography}{99}


\bibitem{hasan} Hasan M Z and Kane C L 2010 \emph{Rev. Mod. Phys.} \textbf{82} 3045

\bibitem{kitrev} Kitagawa T (2012) \emph{Quantum Inf. Proc.} \textbf{11} 1107

\bibitem{asb} Asb\'oth J K, Oroszl\'any L and P\'alyi A 2016 \emph{A Short
    Course
    on Topological Insulators} (Springer: Berlin)

\bibitem{bern} Bernevig B A and Hughes T L (2013) \emph{Topological Insulators
    and Topological Superconductors} (Princeton University Press: Princeton,
    NJ)

\bibitem{stan} Stanescu T D (2017) I \emph{ntroduction to Topological Matter and Quantum Computation} (CRC Press: Boca Raton, FL)

\bibitem{klitz} von Klitzing K, Dorda G and Pepper M (1980)  \emph{Phys. Rev. Lett.}
    \textbf{45} 494

\bibitem{laugh1} Laughlin R B (1981) \emph{Phys. Rev. B} \textbf{23} 5632

\bibitem{thoul} Thouless D J (1983) \emph{Phys. Rev. B} \textbf{27} 6083

\bibitem{tsui} Tsui D C,  Stormer H L and Gossard A C (1982) \emph{Phys. Rev. Lett.} \textbf{48} 1559

\bibitem{laugh2} Laughlin R B (1983) \emph{Phys. Rev. Lett.} \textbf{50} 1395

\bibitem{bouw} Bouwmeester D,  Marzoli I, Karman G P, Schleich W and Woerdman J P (1999) \emph{Phys. Rev. A} \textbf{61} 013410

\bibitem{zhang} Zhang P,  Ren X F, Zou X B, Liu B H,
    Huang Y F and Guo G C (2007) \emph{Phys. Rev. A} \textbf{75} 052310

\bibitem{souto} Souto Ribeiro P H, Walborn S P, Raitz C Jr.,
    Davidovich L and Zagury N (2008) \emph{Phys. Rev. A} \textbf{78} 012326

\bibitem{peretz} Perets H B, Lahini Y, Pozzi F, Sorel M,
    Morandotti R and Silberberg Y (2008) \emph{Phys. Rev. Lett.}
    \textbf{100} 170506

\bibitem{sch} Schreiber A, Cassemiro K N, Poto\v{c}ek V, G\'abris A, Mosley P J, Andersson E, Jex I and Silberhorn C (2010) \emph{Phys. Rev. Lett.}
    \textbf{104} 050502

\bibitem{broome} Broome M A, Fedrizzi A, Lanyon B P, Kassal I,
    Aspuru-Guzik A and  White A G (2010) \emph{Phys. Rev. Lett.} \textbf{104},
    153602

\bibitem{kit1} Kitagawa T, Rudner M S, Berg E and  Demler E (2010) \emph{Phys.
    Rev. A} \textbf{82} 033429

\bibitem{kit2} Kitagawa T, Berg E, Rudner M and Demler E (2010)  
\emph{Phys. Rev. B.} \textbf{82} 235114

\bibitem{kit3}  Kitagawa T, Broome M A, Fedrizzi A, Rudner M S, Berg E,  Kassal I,  Aspuru-Guzik A,  Demler E and White A G (2012) \emph{Nature
    Comm}. {\bf 3} 882


\bibitem{lu1} Lu L, Joannopoulos J D and Solja\v{c}i\'c M (2014) \emph{Nat. Phot.}
    \textbf{8} 821

\bibitem{lu2} Lu L, Joannopoulos J D and Solja\v{c}i\'c M (2016) \emph{Nat. Phys.}
    \textbf{12} 626

\bibitem{cardano} Cardano F, D'Errico A, Dauphin A, Maffei M, Piccirillo B, de Lisio C,  De Filippis G, Cataudella V,  Santamato E, Marrucci L,
    Lewenstein M and Massignan P (2017) \emph{Nat. Comm.} \textbf{8} 15516

\bibitem{simbook} Simon D S (2018) \emph{Tying Light in Knots: Applying Topology to Optics} (Institute of Physics Press/Morgan and Claypool: San Rafael, CA)

\bibitem{moul} Moulieras S, Lewenstein M and Puentes G (2013) \emph{J. Phys. B: At. Mol. Opt. Phys.} {\bf 46} 104005

\bibitem{recht} Rechtsman M K, Lumer Y, Plotnik Y, Perez-Leija A, Szameit A, Segev M (2016) \emph{Optica} {\bf 3} 925

\bibitem{simwave} Simon D S, Osawa S and Sergienko A V (2018) arXiv:1808.10066

\bibitem{sim1} 
Simon D S, Fitzpatrick C A and  Sergienko A V (2016)  \emph{Phys. Rev.  A} \textbf{93} 043845

\bibitem{sim3} Simon D S, Fitzpatrick C A, Osawa S and Sergienko A V (2017)
    \emph{Phys. Rev. A}, \textbf{96} 013858

\bibitem{sim2} Simon D S, Fitzpatrick C A, Osawa S and Sergienko A V (2017)
    \emph{Phys. Rev. A} \textbf{95} 042109

\bibitem{osawa} Osawa S, Simon D S and Sergienko A V (2018) \emph{Opt. Exp.} \textbf{26} 27201

\bibitem{feld1} Feldman E and Hillery M (2004) \emph{Phys. Lett. A} \textbf{324} 277

\bibitem{feld2} Feldman E and Hillery M (2005) \emph{Contemp. Math.} \textbf{381} 71

\bibitem{feld3} Feldman E and Hillery M (2007) \emph{J. Phys. A: Math. Theor.} \textbf{40} 11343

%
%

\bibitem{longhi} Longhi S (2013) \emph{Opt. Lett.} \textbf{38} 3716

\bibitem{ozawa} Ozawa T and Carusotto I, (2014) \emph{Phys. Rev. Lett.}
    \textbf{112} 133902

\bibitem{hafezi} Hafezi M (2014) \emph{Phys. Rev. Lett.} \textbf{112} 210405

\bibitem{taras} Tarasinski B, Asb\'oth J K and Dahlhaus J P (2014)
    \emph{Phys. Rev. A} \textbf{89} 042327

\bibitem{bark} Barkhofen S, Nitsche T, Elster F, Lorz L, G\'abris A, Jex I and Silberhorn C (2016) arXiv: 1606.00299v1 [quant-ph]

\bibitem{maffei} Maffei M, Dauphin A, Cardano F, Lewenstein M and
    Massignan P (2018) \emph{New J. Phys.} \textbf{20} 013023

\bibitem{wan} Wannier G H (1937) 
\emph{Phys. Rev. } \textbf{52} 191

\bibitem{kohn}  Kohn W (1959) 
\emph{Phys. Rev.} \textbf{115} 809

\bibitem{cloiz}  des Cloizeaux J  (1963)  
\emph{Phys. Rev.} \textbf{129} 554

\bibitem{duncan} Duncan C W, \"Ohberg P, Valiente M (2018) \emph{Phys. Rev. B} \textbf{97} 195439 

\bibitem{meyer1} Meyer D A (1997) \emph{Phys. Rev. E} \textbf{55} 5261

\bibitem{meyer2} Meyer D A (1997)  \emph{Int. J. of Mod. Phys.} \textbf{8}  717

\bibitem{meyer3} Meyer D A (1998)  \emph{J. Phys. A: Math. and Gen.} \textbf{31} 2321 



\end{thebibliography}
\end{document}